\documentclass[%
,aps%
 ,twocolumn%
 ,secnumarabic%
 ,superscriptaddress%
,amssymb, amsmath,nobibnotes, aps, prd, floatfix,nofootinbib]{revtex4-2}
\usepackage{docs}%
\usepackage{bm}%
\RequirePackage[utf8]{inputenc}
\RequirePackage{xspace}
\RequirePackage{graphicx}
\RequirePackage{amsmath}
\RequirePackage{amssymb}
 \RequirePackage{xcolor}
 \RequirePackage[sort&compress]{natbib}
 \RequirePackage{hepunits}
 \RequirePackage{enumitem}
 \usepackage{tikz}
 \usepackage[compat=1.1.0]{tikz-feynman}
\usepackage[colorlinks=true,linkcolor=blue]{hyperref}%
 \usepackage{definitions}%
 \usepackage{comments}%
 \usepackage{mathtools}
 \usepackage{slashed}
 \usepackage{mathrsfs}
 \usepackage{changes}
 \usepackage{orcidlink}
 
\expandafter\ifx\csname package@font\endcsname\relax\else
 \expandafter\expandafter
 \expandafter\usepackage
 \expandafter\expandafter
 \expandafter{\csname package@font\endcsname}%
\fi

\begin{document}

\title{Dispersion relations of deeply virtual Compton scattering: investigating twist-4 kinematic power corrections}%
\author{V\'ictor \surname{Mart\'inez-Fern\'andez\orcidlink{0000-0002-0581-7154}}
}%
\email{victor.martinezfernandez@cea.fr}
\affiliation{Irfu, CEA, Université Paris-Saclay, F-91191, Gif-sur-Yvette, France}
\affiliation{Center for Frontiers in Nuclear Science, Stony Brook University, Stony Brook, NY 11794, USA}
\author{Cédric \surname{Mezrag}\orcidlink{0000-0001-8678-4085}}
\email{cedric.mezrag@cea.fr}
\affiliation{Irfu, CEA, Université Paris-Saclay,  F-91191, Gif-sur-Yvette, France}
\date{today}%
\begin{abstract}
  In this paper we include kinematic power corrections up to twist‑four to the deeply virtual Compton scattering dispersion relation.
  We demonstrate that, both for (pseudo‑)scalar and spin‑$1/2$ targets, the formal expression of the $n$‑subtracted leading‑twist dispersion relations is preserved.
  However, the expression of the subtracted constants \emph{is} modified by the kinematic powers.
  Importantly, the minimal-subtracted dispersion relation for the helicity‑conserving amplitude, previously thought to depend only on the Polyakov–Weiss $D$‑term, now also depends on the double distributions $F$ and $K$.
  These results are consistent with the ones obtained previously in the literature.
  Such a mixing may be critical for the Jefferson Lab kinematic range, as it is not suppressed for typical values of $t$ and $Q^{2}$ in the valence region. We therefore expect a strong impact on the attempts to extract pressure forces from DVCS data.
\end{abstract}

\maketitle

\section{Introduction}
The energy momentum tensors (EMTs) of hadrons are today at the core of an intense research activity.
Many theoretical and phenomenological studies have been performed (for instance \cite{Polyakov:2018zvc,Lorce:2018egm,Burkert:2018bqq,Kumericki:2019ddg,Dutrieux:2021nlz,Freese:2021jqs,Dutrieux:2024bgc}).
Lattice and continuum QCD computations have also been performed in the past few years \cite{Alexandrou:2017oeh,Alexandrou:2020sml,Wang:2021vqy,Hackett:2023rif,Nair:2024fit,Yao:2024ixu}.
The goal of this activity is to understand how the macroscopic properties of the nucleon, such as its mass and its spin, emerge from the dynamics of QCD. 

However, connecting the EMT with experimental data is challenging.
Today only an indirect connection, already noticed three decades ago \cite{Ji:1996ek}, is available through generalised parton distributions (GPDs).
Introduced independently in \cite{Mueller:1998fv,Ji:1996ek,Ji:1996nm,Radyushkin:1996ru,Radyushkin:1997ki}, GPDs allow one to describe the amplitude of exclusive processes through factorisation with a coefficient function computed in perturbation theory \cite{Collins:1996fb,Collins:1998be}.
The better studied experimental process connected to GPD is certainly deeply virtual Compton scattering (DVCS) \cite{Ji:1996nm}.
But other exclusive processes are connected to GPDs, such that time-like Compton scattering (TCS) \cite{Berger:2001xd} or deeply virtual meson production (DVMP) \cite{Mueller:2013caa}.
However, all these processes face a severe deconvolution problem \cite{Bertone:2021yyz,Moffat:2023svr}.
Therefore, double DVCS \cite{Belitsky:2003fj,Guidal:2002kt,Deja:2023ahc,Alvarado:2025huq} (DDVCS) and multi-particle exclusive processes have been advocated to bypass this issue \cite{Boussarie:2016qop,Duplancic:2018bum,Grocholski:2021man,Qiu:2022bpq, Qiu:2022pla, Qiu:2023mrm}.
Nevertheless, DVCS remains today the main source of experimental knowledge on GPDs with measurements spanning on the last two decades \cite{Girod:2007aa,HERMES:2012gbh,Jo:2015ema,Defurne:2015kxq,Defurne:2017paw,COMPASS:2018pup,JeffersonLabHallA:2022pnx,CLAS:2022syx}.

This large experimental campaign has triggered many theoretical developments improving the description of DVCS. One can for instance mention the description of the perturbative coefficient function at next-to-next-to-leading order (NNLO) \cite{Braun:2020yib}.
More critical for current facilities running in the valence region, a significant effort has been performed to derive kinematic higher power corrections \cite{Belitsky:2000vx,Braun:2012hq,Braun:2012bg,Braun:2014sta,Braun:2016qlg,Braun:2020zjm,Braun:2022qly}.
These corrections have been recently extended to DDVCS for a (pseudo-)scalar target \cite{Martinez-Fernandez:2025gub}. 
They are expected to contribute up to 40\% of the DVCS amplitude for some of the kinematic area \cite{Defurne:2015kxq}.
As a consequence, they may have a significant impact on the extraction of the pressure and shear forces from experimental data, usually performed through DVCS dispersion relations bounding the real and imaginary part of DVCS amplitude.
DVCS dispersion relations have been derived two decades ago, and the size of NLO corrections \cite{Anikin:2007yh,Dutrieux:2024bgc} have been shown to be of the order of 10\% of the leading contribution \cite{Dutrieux:2024bgc}.
Consequently, there is room for kinematic power corrections to be significantly larger than NLO corrections in the strong coupling constant.

In this paper, we revisit the topic of kinematic power correction up to twist-4 to DVCS dispersion relations for (pseudo-)scalar and spin-$1/2$ targets.
In section \ref{sec:EMT}, we introduce our notations and conventions.
In section \ref{sec:scalar} and \ref{sec:nucleon} we provide a derivation of dispersion relations with higher kinematic-power corrections, for (pseudo-)scalar and spin-$1/2$ targets respectively, adapting the proof of Ref.~\cite{Dutrieux:2024bgc}.
In the nucleon case, an expression of the DVCS subtraction constant including kinematic twist-four corrections was derived and expressed as a double integral over GPDs in Ref.~\cite{Braun:2014sta}.
  Our new expression relying on a single integral over double distributions~\cite{Mueller:1998fv,Radyushkin:1997ki} agrees with this pioneering result.
 Beyond the DVCS subtraction constant, we also derived the twist-four corrections to higher-subtracted dispersion relations connecting the Mellin moments of the imaginary part of the CFFs with double distributions.
In section~\ref{sec:Deconvolution}, we compute the impact of the kinematic corrections to the Gegenbauer modes of the Polyakov-Weiss $D$-term~\cite{Polyakov:1999gs} and highlight that these corrections may help one to extract these modes from experimental data.
Then, in section~\ref{sec:Numerical} we provide numerical estimates, demonstrating the relevance of power-corrections.
Finally, we conclude in section \ref{sec:conclusion}.


\section{Accessing the energy-momentum tensor via generalised parton distributions}
\label{sec:EMT}

The energy momentum tensor (EMT) of a hadron is obtained by projecting the local and gauge invariant operator $\texttt{T}^{\mu\nu}$ between two hadron states off-diagonal in momentum.
The momentum transfer is labelled $\Delta = p'-p$, introducing also $t=\Delta^2$ and the average momentum $\bp = (p+p')/2$.
For a spin-0 target, the non-conserved, symmetric tensor decomposition involves three form factors \cite{Bakker:2004ib,Leader:2013jra}:
\begin{align}
  \bra{p'} \texttt{T}^{\mu\nu}_a(0)\ket{p} = & 2 \bp^\mu \bp^\nu A_a(t) + 2\left(\Delta^\mu\Delta^\nu - \eta^{\mu\nu}\Delta^2 \right)C_a(t) \nonumber \\
  & + \eta^{\mu\nu} M^2 \bar{C}_a(t)  
\end{align}
where the index $a$ labels quark flavours or gluon contributions.
For a spin-$1/2$ hadron, five form factors are required to parameterise the non-conserved, asymmetric EMT:
\begin{align}
  &\bra{p', s'} \texttt{T}^{\mu\nu}_a(0)\ket{p, s}
    = \\
  &\hspace{15pt}\bar u(p', s') \Bigg\{
    \frac{\bp^\mu \bp^\nu}{M}\,A_a(t)
    + M \eta^{\mu\nu}\bar C_a(t) \nonumber \\
  &\hspace{15pt}+ \frac{\Delta^\mu\Delta^\nu - \eta^{\mu\nu}\Delta^2}{M}\, C_a(t)+ \frac{\bp^{[\mu} i\sigma^{\nu]\rho}\Delta_\rho}{4M}\,D_a(t)
    \nonumber
  \\
  &\hspace{15pt}+ \frac{\bp^{\{\mu} i\sigma^{\nu\}\rho}\Delta_\rho}{4M}\left[A_a(t)+B_a(t)\right]\Bigg\} u(p, s) \,.
\end{align}
One can build a physical interpretation for several of these form factors in terms of ``mechanical properties'' of hadrons, like internal pressure or shear forces distributions \cite{Polyakov:2002yz,Polyakov:2018zvc,Lorce:2018egm}, and their decomposition in terms of quarks and gluons.
  
The form factors of the EMT cannot be probed directly experimentally.
However, some of them are connected with the generalised parton distributions (GPDs) \cite{Mueller:1998fv,Ji:1996ek,Ji:1996nm,Radyushkin:1996ru,Radyushkin:1997ki} through their Mellin moments over $x$, the average longitudinal light-front momentum fraction of the active parton (we use the conventions of Ref.~\cite{Diehl:2003ny}):
\begin{align}
  \label{eq:mel1H}
  \int_{-1}^{1}\ud x\,x\,H^q(x,\xi,t)&=A_q(t)+4\xi^2C_q(t) \,, \\
  \label{eq:mel1E}
  \int_{-1}^{1}\ud x\,x\,E^q(x,\xi,t)&=B_q(t)-4\xi^2C_q(t) \,,
\end{align}
where Eq.~\eqref{eq:mel1E} is relevant for spin-$1/2$ hadrons. The last kinematic variable, the skewness $\xi$ is defined in terms of the ``plus'' light-components of the initial- and final-state momenta of the hadron as
		\begin{equation}
			\xi = -\frac{\Delta^+}{2\bar{p}^+} = \frac{p^+ - p'^+}{p^+ + p'^+}\,.
		\end{equation}
		In light-cone coordinates we can parameterise any four-vector $v$ as
		\begin{equation}
			v^\mu = v^+n'^\mu + v^-n^\mu + v_\perp^\mu\,,
		\end{equation}
		with the conditions $n^2=n'^2=0$, $nn'\neq 0$ and $nv_\perp = n'v_\perp = 0$, such that $v^+ = vn/(nn')$ and $v^- = vn'/(nn')$. Taking into account that in DVCS $q'^2=0$, then we can select
		\begin{align}
			n^\mu & = q'^\mu \,, \label{n} \\
			n'^\mu & = -q^\mu + \left( 1-\frac{t}{\scale^2} \right)q'^\mu\, \mbox{ with }\scale^2 = Q^2+t\,. \label{nPrime}
		\end{align}
		This choice, introduced in Ref.~\cite{Braun:2020zjm}, is of physical importance as it keeps the leading kinematic contribution ({\it leading twist}, LT) $U(1)$-electromagnetic gauge invariant. In fact, the Compton tensor, encoding the partonic distributions by Compton form factors (CFFs), is dominated at LT by the transverse metric and the CFF labelled $\cffH^{++}$,
		\begin{equation}
			T^{\mu\nu}_{\rm LT} = -g_\perp^{\mu\nu}\cffH^{++} + \cdots\,.
		\end{equation}
		Therefore, if the in or out photon carry a transverse component, the LT approximation is $U(1)$-violating. For this reason, we span the longitudinal plane by the momenta $q,q'$.
Through GPDs, some of the form factors of the EMT can thus indirectly be probed experimentally. 

We also recall that GPDs can be written in terms of Double Distributions $F$ and $K$ \cite{Mueller:1998fv,Radyushkin:1997ki} (see for instance ref. \cite{Teryaev:2001qm,Chouika:2017dhe,Chouika:2017rzs,DallOlio:2024vjv} for the interpretation of DDs in terms of Radon transforms) plus the so-called Polyakov-Weiss $D$-term \cite{Polyakov:1999gs}:
\begin{align}
  \label{eq:DDF}
  H^q(x,\xi,t) = & \int_\Omega \textrm{d}\beta \textrm{d}\alpha \bigg[F^q(\beta,\alpha,t) \nonumber \\
    &\hspace{-20pt} +\xi D^q(\alpha,t)\delta(\beta) \bigg] \times \delta(x-\beta-\alpha\xi),\\
    \label{eq:DDK}
  E^q(x,\xi,t) = & \int_\Omega \textrm{d}\beta \textrm{d}\alpha \bigg[ K^q(\beta,\alpha,t)\nonumber \\
&\hspace{-20pt}-\xi D^q(\alpha,t)\delta(\beta) \bigg]  \times \delta(x-\beta-\alpha\xi),
\end{align}
where $\Omega = \{(\alpha,\beta)| |\alpha|+|\beta|\le 1\}$.
For convenience, we already introduce the so-called magnetic Double Distribution given as \cite{Belitsky:2005qn,Radyushkin:2013hca,Mezrag:2013mya}:
\begin{equation}
  \label{eq:MagneticDD}
  N^q(\beta,\alpha,t) = \frac{F^q(\beta,\alpha,t) + K^q(\beta,\alpha,t)}{2}
\end{equation}
Note that the first Mellin moment of the $D$-term yields the Form Factor $C_a(t)$ in Eq.~\eqref{eq:mel1H}:
\begin{align}
    \label{eq:eihwjcnalkxm}
  C_q(t) = \frac{1}{4}\int_{-1}^{1}\ud \alpha\,\alpha D^q(\alpha,t)\,.
\end{align}

It has been argued in the past that $C_q(t)$ could have been extracted using deeply virtual Compton Scattering (DVCS) dispersion relation, bypassing the deconvolution of GPDs which reveals itself at best delicate \cite{Bertone:2021yyz,Moffat:2023svr}.
Indeed, the Compton form factors ($\mathcal{H}$, $\mathcal{E}$\dots ) parameterising the DVCS amplitude are related to GPDs through the convolution:
\begin{align}
  \label{eq:CFFquarkDef}
  \mathcal{H}^q(\xi,t,Q^2) &= \int_{-1}^1 \frac{\textrm{d}x}{\xi} T^q\left(\frac{x}{\xi};\frac{Q^2}{\mu^2},\alpha_s ,\frac{t}{\mathbb{Q}^2}\right) H^q(x,\xi,t,\mu^2),
\end{align}
where $T^q$ represents here the DVCS coefficient function, computed in perturbation theory, and $\mathbb{Q}^2 = Q^2 + t$.
Indeed, from kinematic higher-twist studies of the DVCS amplitude such as \cite{Braun:2022qly,Braun:2025xlp,Martinez-Fernandez:2025gub}, the natural scale for expansion on twist is $\mathbb{Q}^2$ rather than $Q^2$. The difference between both of them is a higher-twist term so one can choose either one. In this work, we select $\mathbb{Q}^2$ as it simplifies expansions and provides a direct comparison with previous literature. 

However, as already pointed out in \cite{Dutrieux:2021nlz,Dutrieux:2024bgc}, a dispersive approach \cite{Teryaev:2005uj,Anikin:2007yh,Diehl:2007jb,Dutrieux:2024bgc} does not preclude facing an ill-posed deconvolution problem.
In that regard, kinematic higher-twist corrections provide a new lever arm, adding an explicit $t$-dependence in the coefficient function that comes with a more involved $x$ behaviour (typically involving $\textrm{Li}_2$ functions). 
However, such corrections will also impact the derivation of the dispersion relations.
Consequently, in the following we re-derive the dispersion relations, taking into account the first power corrections.


\section{Power corrections to dispersion relations: the scalar case}
\label{sec:scalar}
In this section we adapt the proof of Ref.~\cite{Dutrieux:2024bgc} to take into account the $t$ and target-mass corrections. 

\subsection{DVCS Dispersion relation with power corrections for $\mathcal{H}^{ij}$ amplitude}
\label{sec:DRproof}

We start by considering DVCS on a scalar target (such as a pion in a Sullivan process \cite{Amrath:2008vx,Chavez:2021llq,Chavez:2021koz,Castro:2025rpx}, or the ${}^4\textrm{He}$ nucleus):
\begin{equation}
  \label{eq:comptonScattering}
    N(p)+\gamma^{(*)}(q)\to N(p')+\gamma(q')\,.
\end{equation}
For a process of this kind: 
\begin{equation}
    p^2=p'^2=M^2\,,\qquad q^2=-Q^2\,,\qquad q'^2 = 0,
\end{equation}
where $M$ is the mass of the hadron and, from now on,  $Q^2$ is positive and $t$ is negative. Then, the Mandelstam variables take the form
\begin{align}
    s & = M^2-Q^2+2p\cdot q\,,\\
    t & = -2p\cdot \Delta\,,\\
    u & = M^2-2p\cdot q-t\,,\\
    s+t+u & = 2M^2-Q^2\,. \label{s+u}
\end{align}
For fixed negative values of $t$, the process can be described with just one variable: $s$, $u$ or an appropriate combination of both of them. Following Ref.~\cite{Dutrieux:2024bgc}, one can choose (Note the relative minus sign in the definition of $\nu$ in Eq.~(\ref{eq:nu}) here with respect to Ref.~\cite{Dutrieux:2024bgc}):
\begin{equation}
  \label{eq:nu}
    \nu = -\frac{s-u}{s+u}\,.
\end{equation}
This choice is particularly convenient as it is the inverse of the DVCS skewness $\xi$ at leading twist (LT) accuracy. However, higher power corrections modify this simple relation as we will see below. 
Reshuffling the expression connecting the Mandelstam variables, one gets:
\begin{align}
  \label{eq:s+u_dvcs}
  s+u & = 2M^2-Q^2+|t| = -Q^2\left[ 1 - \frac{|t|+2M^2}{Q^2} \right] \nonumber\\
      & = -(Q^2+t)\left[ 1-\frac{2M^2}{Q^2+t} \right]\, \nonumber \\
  & = - \mathbb{Q}^2\left[ 1-\frac{2M^2}{\mathbb{Q}^2} \right]\, .
\end{align}
where $s+u<0$ can be kept as long as:
\begin{equation}
  \label{eq:DRCondition}
  M^2/(Q^2+t) < 1/2 \quad \textrm{and}\quad  |t| < Q^2.
\end{equation}
In such a case, there exists a region of the kinematics domain for which both $s$ and $u$ are negative, hence closed for particle production.
As we will demonstrate in the following, this region corresponds to $-1 < \nu < 1$ (see Eq.~\eqref{eq:domNu}).
The amplitude is thus real and analytic on a segment of the real axis allowing us to define analytic continuation in the entire complex plane through the Schwartz principle (see Ref.~\cite{Dutrieux:2024bgc} for details).

Note also that:
\begin{equation}
  \label{eq:s-u_dvcs}
    s-u = -Q^2+t+4pq = 4\bp q = 4\bp q' \,,
\end{equation}
with $\bp = (p+p')/2$, and  
\begin{equation}
  \label{eq:Deltaq'}
  -\Delta q' = -qq' = \frac{1}{2} \left[ \underbrace{(q-q')^2}_{t} - q^2 - \underbrace{q'^2}_{0} \right] = \frac{Q^2+t}{2}\,.
\end{equation}
Taking into account the choice of longitudinal plane given in Eqs.~(\ref{n}) and (\ref{nPrime}), the skewness reads
\begin{equation}
  \xi = -\frac{\Delta n}{2\bp n} = -\frac{\Delta q'}{2\bp q'} = \frac{t+Q^2}{s-u} = \frac{\mathbb{Q}^2}{s-u}\,.
\end{equation}
Using this expression for $\xi$ together with the hard scale $\scale^2=-2qq'=Q^2+t$ and Eq.~\eqref{eq:s+u_dvcs}, $\nu$ is given by
\begin{equation}
  \label{eq:nu(xi)}
    \nu = \frac{1/\xi}{1-2M^2/\scale^2} \xrightarrow[\textrm{limit}]{\textrm{Bjorken}}\frac{1}{\xi}\,.
  \end{equation}
Since for $M^2/\scale^2 \ge 1/2$ there is at least one channel open for particle production, we will consider $M^2/\scale^2 < 1/2$ in this work. Note, however, that this is a sufficient but not necessary condition.

Because $\nu\ge 1$ implies $s\ge 0$, the amplitude will be analytically continued to the upper half of the complex plane of $s$: $s\to s+i\eta$ while $u\to u-i\eta$ with $\eta\in\R^+$. Then, $\nu\to\nu+i\eta$. For the case of $\nu\le -1$, we have $u\ge 0$, then $s\to s-i\eta$ while $u\to u+i\eta$, and $\nu\to\nu-i\eta$. This implies that the amplitude is recovered for $\nu>0$ by approaching the real axis from above, while for $\nu<0$ it is approached from below. For $\nu_0\in\R^+$:
\begin{align}
    \mathcal{F}(\nu_0) & = \lim_{\eta\to 0^+}\mathcal{F}(\nu_0+i\eta)\,,\\
    \mathcal{F}(-\nu_0) & = \lim_{\eta\to 0^+}\mathcal{F}(-\nu_0-i\eta)\,,
\end{align}
and by Schwartz's reflection principle:
\begin{align}
    \mathcal{F}(\nu_0-i\eta) & = \mathcal{F}^*(\nu_0+i\eta)\,,\\
    \mathcal{F}(-\nu_0+i\eta) & = \mathcal{F}^*(-\nu_0-i\eta)\,.
\end{align}
Because $\nu$ is inversely proportional to $\xi$, cf.~Eq.~(\ref{eq:nu(xi)}), $\mathcal{F}(\nu\pm i\eta)$ is equivalent to $\mathcal{\widetilde{F}}(\xi\mp i\eta)$ as $\eta\to 0^+$.

\begin{figure}
    \centering
    \includegraphics[width=0.5\linewidth]{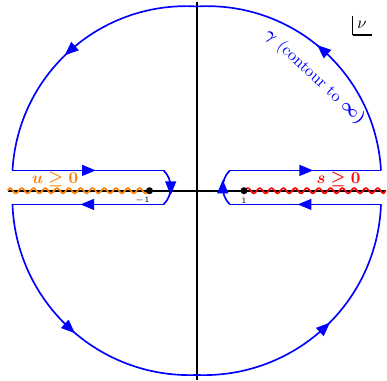}
    \caption{Complex plane for variable $\nu$ where the region where analyticity is not granted, i.e.~the physical domain $\nu\in(-\infty,-1]\cup[1,\infty)$, has been highlighted in red for the positive-$s$ region and in orange for the positive-$u$ segment. Note that they are exchanged with respect to Ref.~\cite{Dutrieux:2024bgc}. The contour $\gamma$ runs over and its interior is within the analytic domain of $\nu$ so that $\oint_\gamma d\nu'\ \frac{\mathcal{F}(\nu')}{\nu'-\nu} = 0$ for $\nu$ in the physical region, accordingly to Cauchy's integral theorem. }
    \label{fig:analyticityInNu}
\end{figure}

\begin{figure}
    \centering
    \includegraphics[width=0.5\linewidth]{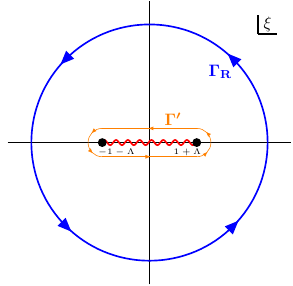}
        \caption{Complex plane for variable $\xi$ where the region where analyticity is not granted, i.e.~$\xi\in[-1-\Lambda,1+\Lambda]$, has been highlighted in red. Note that this interval is larger than the physical domain which corresponds to $|\xi| < 1$.}
    \label{fig:analyticityInXi}
\end{figure}

Finally, $\mathcal{F}(\nu)$ is granted to be analytic for 
\begin{equation}
  \label{eq:domNu}
    \nu\in \left\{ \C - (-\infty,-1]\cup[1,+\infty) \right\}\,,
\end{equation}
as illustrated in Fig.~\ref{fig:analyticityInNu}, while $\mathcal{\widetilde{F}}(\xi)$ for
\begin{equation}
  \label{eq:domXi}
  \xi\in \left\{ \C - [-1-\Lambda,1+\Lambda] \right\}\,,\qquad \Lambda = \frac{2M^2}{\scale^2}\,,
\end{equation}
as shown in Fig.~\ref{fig:analyticityInXi}.
The factor $\Lambda$ comes from
\begin{equation}
    |\nu| < 1 \Rightarrow |\xi| > \frac{1}{1-2M^2/\scale^2} = 1+\frac{2M^2}{\scale^2} + O\left(\frac{M^4}{\scale^4}\right) > 1\,.
\end{equation}
This expansion is possible thanks to the condition $M^2/\scale^2<1/2$. Note the difference with respect to a standard mass correction which comes in powers of  $\xi M/\scale$, cf.~\cite{Braun:2022qly, Martinez-Fernandez:2025gub}. There, the skewness dependence guarantees that the mass corrections are tamed for large nuclei. Indeed, up to $t/\scale^2$ terms and binding energy corrections ($M\approx A M^{\rm proton}$, with $A$ the mass number), the skewness takes the form
	\begin{equation}
		\xi\approx \frac{x_B}{2-x_B} \approx \frac{x_B^{\rm proton}}{2A-x_B^{\rm proton}} \,,
	\end{equation}
	thus the mass correction reads
	\begin{equation}
		\frac{\xi M}{\scale} = \frac{x_B^{\rm proton}}{2-x_B^{\rm proton}/A}\frac{M^{\rm proton}}{\scale} \xrightarrow[A\to\infty]{} \frac{x_B^{\rm proton}M^{\rm proton}}{2\scale}  \,.
	\end{equation}
	This implies that the mass corrections, even for large nuclei, ought to be of the same order as for the proton. However, the condition for granted analyticity ($M^2/\scale^2<1/2$) is not regulated by the skewness, making it dependent of the target mass.

From the considerations above, $\mathcal{F}(\nu)$ can be written around $\nu=0$ as a entire series of radius of convergence $|\nu| <1 $ in the complex plane as:
  \begin{align}
    \label{eq:FnuPower}
    \left. \mathcal{F}(\nu)\right|_{|\nu|< 1} = \sum_{j=0}^\infty f_j \nu^j
  \end{align}
  where the $f_j$ are guaranteed to be real because of the Schwartz principle.
  Consequently, within the domain~\eqref{eq:domXi}, $\mathcal{\widetilde{F}}(\xi)$ can be expanded such that
\begin{equation}
    \left.\mathcal{\widetilde{F}}(\xi)\right|_{|\xi|>1+\Lambda} = \sum_{j=0}^\infty f_j\frac{1}{\xi^j}\,,\qquad f_j\in\R\,.
\end{equation}

Going back to standard CFF notations, and regardless of the incoming ($A$) or outgoing ($B$) photon polarisation, the CFF $\cffH^{AB}$ can thus be expanded as:
\begin{equation}
  \label{eq:cffH_seriesXi}
    \left. \cffH^{AB}(\xi)\right|_{|\xi|>1+\Lambda} = \sum_{j=0}^\infty h_j^{AB}\frac{1}{\xi^j}\,,\qquad h_j^{AB}\in\R\, .
\end{equation}
At this point, the demonstration follows the one of Ref.~\cite{Dutrieux:2024bgc}. Briefly we introduce the $n$-subtracted integral $\mathcal{I}_n^{AB}(\xi)$  over the contour $\Gamma_\textrm{R}$ of Fig.~\ref{fig:analyticityInXi}:
\begin{align}
  \label{eq:InXi_GammaR}
  \left. \mathcal{I}_n^{AB}(\xi)\right|_{|\xi|\in \textrm{branch cut}} =& \oint_{\Gamma_{\rm R}}d\xi'\ \frac{\cffH^{AB}(\xi')}{\xi'-\xi}\left(\frac{\xi'}{\xi}\right)^n\, \nonumber \\
                             = &  2\pi i\sum_{j=0}^n h_j^{AB}\frac{1}{\xi^j}\,  
\end{align}
which truncates the series of Eq.~\eqref{eq:cffH_seriesXi} for $\xi$ in the physical region.
This result, obtained with the residue theorem, has to match that of the same integration with respect to the closed curve $\Gamma'$ because both $\Gamma_\textrm{R}$ and $\Gamma'$ are inside the domain~\eqref{eq:domXi} and, therefore, are homotopic.

Integrating over $\Gamma'$ while keeping $\xi$ on the branch cut, one is left with:
\begin{align}
  \label{eq:Gamma'Integration}
  &\mathcal{I}_n^{AB}(\xi) \nonumber \\
    = &\int_{-(1+\Lambda)}^{1+\Lambda}d\xi'\ \frac{\cffH^{AB}(\xi'-i0)}{\xi'-\xi-i0}\left(\frac{\xi'-i0}{\xi}\right)^n - \textrm{(c.c.)} \nonumber\\
  = &\textrm{PV}\int_{-(1+\Lambda)}^{1+\Lambda}d\xi'\ \frac{\cffH^{AB}(\xi'-i0)}{\xi'-\xi}\left(\frac{\xi'-i0}{\xi}\right)^n \nonumber \\
  & \quad + i\pi\cffH^{AB}(\xi-i0)\left(1-\frac{i0}{\xi}\right)^n - \textrm{(c.c.)} \nonumber\\
  = & \textrm{PV}\int_{-(1+\Lambda)}^{1+\Lambda}d\xi'\ \frac{\cffH^{AB}(\xi'-i0)-\cffH^{AB}(\xi'+i0)}{\xi'-\xi}\left(\frac{\xi'}{\xi}\right)^n \nonumber \\
  & \quad + i\pi\left[ \cffH^{AB}(\xi-i0) + \cffH^{AB}(\xi+i0) \right]\,.
\end{align}
where (c.c.) stands for ``complex conjugate'' and PV represents Cauchy's principal value. From the first to the second line we used the Sokhotski-Plemelj formula. Considering $\xi\in (0,1)$, the singular behaviour around zero skewness is avoided and we can safely take $i0\cffH^{AB}(\xi'\pm i0)\to 0$ when going from the second to the third line. Remember that $\cffH^{AB}(\xi)\sim \xi^{-\alpha}$ as $\xi\to 0$, then the formula above is only valid for $n > \alpha-1$.

One finally gets for physical values of $\xi$:
\begin{align}
  \label{eq:DR_withLambda}
  \sum_{j=0}^n h_j^{AB}\frac{1}{\xi^j} & = \textrm{Re}(\cffH^{AB}(\xi))\, \nonumber \\
  &\phantom{=} + \frac{1}{\pi}\textrm{PV}\int_{-(1+\Lambda)}^{1+\Lambda} d\xi'\ \frac{\textrm{Im}(\cffH^{AB}(\xi'))}{\xi'-\xi}\left(\frac{\xi'}{\xi}\right)^n\, .
\end{align}
This expression coincides with the one obtained in Ref.~\cite{Dutrieux:2024bgc} up to the term $\Lambda$ in the integration limits.
However, according to the factorisation theorem, the Compton form factor (CFF) of an exclusive process such as DVCS is given by the convolution of a coefficient function $T$ with a generalised parton distribution (GPD) (see Eq.~\eqref{eq:CFFquarkDef}).
The imaginary part of the CFF $\cffH $ is generated solely from the so-called DGLAP kinematic region
of the GPDs, \ie $|x|\ge |\xi|$.
For $|\xi|\ge 1$, the convolution of Eq.~\eqref{eq:CFFquarkDef} does not probe the DGLAP region and thus, the imaginary part of the CFF vanishes.

In the ERBL region ($|x|<|\xi|$), no imaginary part is generated because of the different signs of the momentum fractions $x+\xi$ and $x-\xi$. In that situation, no on-shell intermediate state can be generated in the $u$ or $s$ channels.
Indeed, the imaginary part of the amplitude, according to Cutkosky's rules, is generated by on-shell intermediate states (poles or cuts), in the $s$ and $u$ channels, respectively (we recall that $t$ is chosen space-like from the start).

Following this argument, one recovers the same result than Ref.~\cite{Dutrieux:2024bgc} for physical values of $\xi$:
\begin{align}
  \label{eq:DR_Standard}
  \sum_{j=0}^n h_j^{AB}\frac{1}{\xi^j} & = \textrm{Re}(\cffH^{AB}(\xi))\, \nonumber \\
  & \phantom{=} + \frac{1}{\pi}\textrm{PV}\int_{-1}^{1} d\xi'\ \frac{\textrm{Im}(\cffH^{AB}(\xi'))}{\xi'-\xi}\left(\frac{\xi'}{\xi}\right)^n\, .
\end{align}

\subsection{Subtraction constant with higher power corrections}
\label{sec:Scalarh0}

The formal expression of the dispersion relation given in terms of the real and imaginary part of the CFF is left unchanged compared to the leading twist case.
However, the way the $h^{AB}_j$ are connected with the GPDs is significantly impacted, as the expression of the coefficient function is strongly modified.
We explore here the impact of these corrections and their dependence for the $\cffH^{++}$ Compton form factor, of great importance for being the only one participating of the LT amplitude. 
The connection between $\cffH^{++}$ and the GPDs can be written in a more compact way, separating terms by their number of derivatives $\dxi$:
\begin{align}
  \label{eq:cffH_full_compact}
  \cffH^{++} = & \int_{-1}^1\frac{\textrm{d}x}{\xi}\Bigg\{ T_0^{++}\left(\frac{x}{\xi},\frac{t}{\scale^2}\right) H \nonumber \\
                 & \quad + \frac{t}{\scale^2}\xi\dxi\left( T_1^{++}\left(\frac{x}{\xi}\right) H \right) \nonumber \\
               & \quad + \frac{-2\xi^2\bp_\perp^2}{\scale^2}\xi^2\dxi^2\left(T_1^{++}\left(\frac{x}{\xi}\right) H\right) \Bigg\} \nonumber \\
  & \quad + O(\textrm{tw-6},\, \alpha_s\cdot\textrm{tw-4})\,,
\end{align}
where
\begin{align}
  \label{eq:T0++_and_T1++}
  T_0^{++}\left(\frac{x}{\xi}, \frac{t}{\scale^2}\right) = &C_{\rm LT}^{(+)}\left(\frac{x}{\xi}\right) +  \frac{t}{\scale^2}\wpbbiii^{(+)}\left(\frac{x}{\xi}\right) \nonumber \\
  & - \frac{t}{\scale^2}\frac{\mathcal{L}^{(+)}\left(\frac{x}{\xi}\right)+C_0^{(+)}\left(\frac{x}{\xi}\right)}{2} \,,\\
    T_1^{++}\left(\frac{x}{\xi}\right) & = \frac{\mathcal{L}^{(+)}\left(\frac{x}{\xi}\right) - \wpbbiii^{(+)}\left(\frac{x}{\xi}\right)}{2}\,.
\end{align}
The reader is invited to refer to appendix~\ref{app:coefficientfunction} for complete definitions of all previous functions.
Here we just highlight that $C_{\rm LT}^{(+)}$ is the leading-twist coefficient function (at arbitrary precision in $\alpha_s$), while $\mathcal{L}^{(+)}$ and $\wpbbiii^{(+)}$ arise from kinematic higher-twist corrections. 
Considering implicitly the $t$-dependence, one can inject the DDs representation \eqref{eq:DDF} term by term in the previous equation.
Doing so allows us to analytically continue the GPDs for $|\xi|>1$ as in this region the Radon transform \eqref{eq:DDF} admits non-vanishing values for $x<|\xi|$ instead of $|x|<1$.
This is a consequence of the support of the Double Distributions $|\beta|+|\alpha| \le 1$. 
We treat the terms one by one following the decomposition:
  \begin{align}
    \label{eq:H++decomposition}
    \cffH^{++} = & \cffH^{++}_0 + \cffH^{++}_1 + \cffH_2^{++}
  \end{align}
  starting by analytically continuing $ \cffH^{++}_0$ for $|\xi| \ge 1$ with\footnote{
    Following our argument before eq. \eqref{eq:DR_Standard}, the natural bound for $|\xi|$ is $|\xi|>1+\Lambda$. However, once we factorise the amplitude into a GPD and a coefficient functions, the analytic properties of the coefficient function dictate the one of the amplitudes. Thus, the natural bound obtained at the level of the amplitude can be relaxed once we assume the factorisation regime holds. As we mentioned before, $s$ and $u$ negative is a sufficient condition to guarantee that the amplitude is analytic, not a necessary one.
  }:
\begin{align}
  \label{eq:cffH_0}
    \cffH^{++}_0  = &\int_{-\xi}^\xi dx\ \frac{1}{\xi} T_0^{++}(x/\xi, t/\scale^2) H \nonumber\\
                  = & \intba\int_{-\xi}^\xi\frac{\textrm{d}x}{\xi}T_0^{++}\left(\frac{x}{\xi},\frac{t}{\scale^2}\right) \nonumber \\
  & \times \deltaxba \left[\Fba+\xi\Da\delta(\beta)\right] \nonumber\\
  = & \intba \frac{1}{\xi}T_0^{++}\left(\frac{\beta}{\xi}+\alpha, \frac{t}{\scale^2}\right) \Fba  \nonumber \\
  & + \int_{-1}^1\textrm{d}\alpha\ T^{++}_0\left(\alpha, \frac{t}{\scale^2}\right)D(\alpha) \nonumber\\
  = & \sum_{n=0}^\infty\frac{1}{n!}\intba \frac{\beta^n}{\xi^{n+1}}T_0^{++\,(n)}(\alpha, t/\scale^2) \Fba \nonumber \\
  & + \int_{-1}^1d\alpha\ T^{++}_0(\alpha, t/\scale^2)D(\alpha)\, ,
\end{align}
where we use the notation $f^{(n)}(y) = \left.\frac{\partial^n f(x)}{\partial x^n}\right|_{x=y}$\,. From the 2nd to the 3rd line, we integrate with respect to $x\in(-\xi,\xi)$ with the $\deltaxba$. From the 3rd to the 4th line, we expand the coefficient function in powers of $1/\xi$ which is only possible in the unphysical domain of $\xi$. This is precisely what we want to do in order to identify the different coefficients $h^{++}_j$ from series~(\ref{eq:cffH_seriesXi}), only valid when the amplitude is factorised for $|\xi|>1$, and be able to read out the subtraction constant ($h_0^{++}$). This is the strategy followed in Ref.~\cite{Dutrieux:2024bgc} and here.

The term on $\dxi$ is given by (again for $|\xi|>1$):
\begin{align}
  \label{eq:cffH_1}
  \cffH^{++}_1  = &\frac{t}{\scale^2}\int_{-\xi}^{\xi}dx\ \dxi\left[ T_1^{++}(x/\xi) H \right] \nonumber\\
  = & \frac{t}{\scale^2}\dxi\left[ \intba T_1^{++}\left(\frac{\beta}{\xi}+\alpha\right)\Fba \right. \nonumber \\
                  & + \left. \xi\int_{-1}^1 \textrm{d}\alpha\ T_1^{++}(\alpha)\Da \right]  \nonumber\\
  = & \frac{t}{\scale^2}\left[\sum_{n=0}^\infty \frac{-n}{n!}\intba \frac{\beta^n}{\xi^{n+1}}T^{++\,(n)}_1(\alpha)\Fba \right. \nonumber \\
  & \left. + \int_{-1}^1 d\alpha\ T^{++}_1(\alpha)\Da\right] \nonumber\\
  = &\frac{t}{\scale^2}\left[\sum_{n=0}^\infty \frac{-1}{n!}\intba \frac{\beta^{n+1}}{\xi^{n+2}}T_1^{++\,(n+1)}(\alpha)\Fba \right. \nonumber \\
  & \left. + \int_{-1}^1\textrm{d}\alpha\ T_1^{++}(\alpha)D(\alpha)\right]\,
\end{align}
while the term with $\dxi^2$ in the region $|\xi|>1$ is
\begin{align}
  \label{eq:cffH_2}
  \cffH_2^{++}  = &\frac{-2\xi^3\bp_\perp^2}{\scale^2}\int_{-\xi}^{\xi}dx\ \dxi^2\left[ T_1^{++}(x/\xi) H \right] \nonumber\\
  = &\frac{-2\xi^2\bp_\perp^2}{\scale^2} \dxi^2\left[ \sum_{n=0}^\infty \frac{1}{n!}\intba \frac{\beta^n}{\xi^n}T_1^{++\,(n)}(\alpha)\Fba \right. \nonumber \\
                  & \left. + \xi\int_{-1}^1d\alpha\ T_1^{++}(\alpha)\Da \right] \nonumber\\
  = &\frac{-2\xi^3\bp_\perp^2}{\scale^2} \sum_{n=0}^\infty \frac{n(n+1)}{n!}  \intba \frac{\beta^nT_1^{++\,(n)}(\alpha)}{\xi^{n+2}}\Fba \nonumber\\
  = &\frac{-2\xi^3\bp_\perp^2}{\scale^2} \sum_{n=0}^\infty \frac{n+2}{n!}  \intba \frac{\beta^{n+1}T_1^{++\,(n+1)}(\alpha)}{\xi^{n+3}}\Fba \nonumber\\
  = &\frac{-2\bp_\perp^2}{\scale^2} \sum_{n=0}^\infty \frac{n+2}{n!}  \intba \frac{\beta^{n+1}T_1^{++\,(n+1)}(\alpha)}{\xi^{n}}\Fba \,.
\end{align}
Taking into account that $\bp_\perp^2$ depends on $\xi$:
\begin{equation}
  \label{eq:bpPerp2}
  \bp_\perp^2 = M^2 - \frac{t}{4}\left(1-\frac{1}{\xi^2} \right)\, ,
\end{equation}
 as well as the above series $\cffH^{++}_i$, we get for $|\xi| >1$:
\begin{widetext}
  \begin{align}
    \sum_{j=0}^\infty h_j^{++}\frac{1}{\xi^j} = & \int_{-1}^1d\alpha\ \left[ T_0^{++}(\alpha,t/\scale^2) + \frac{t}{\scale^2}T_1^{++}(\alpha) \right]\Da - \frac{4M^2-t}{\scale^2}\intba \beta\Fba T_1^{++\,(1)}(\alpha) \nonumber\\
                                                & + \frac{1}{\xi}\intba \Fba\left[ T^{++}_0(\alpha,t/\scale^2) - \frac{6M^2-3t/2}{\scale^2}\beta^2 T_1^{++\,(2)}(\alpha) \right] \nonumber\\
                                                & + \sum_{n=2}^\infty \frac{1}{\xi^n} \intba \Fba\Bigg[ \beta^{n-1} \left\{ \frac{T_0^{++\,(n-1)}(\alpha,t/\scale^2)}{(n-1)!} - \frac{t}{\scale^2}\frac{n+2}{2\cdot(n-2)!} T_1^{++\,(n-1)}(\alpha) \right\} \nonumber\\
                                                & \phantom{+ \sum_{n=2}^\infty \frac{1}{\xi^n} \intba \Fba\Bigg[} - \beta^{n+1}\frac{2M^2-t/2}{\scale^2}\frac{n+2}{n!}T_1^{++\,(n+1)}(\alpha)  \Bigg]\,.
  \end{align}
\end{widetext}
The double distribution $\Fba$ is even in $\alpha$. Taking into account that $\int_{\Omega} \textrm{d}\beta \textrm{d}\alpha $ is done for a symmetric interval in both $\beta$ and $\alpha$, the terms multiplying this DD and that are odd in $\alpha$ vanish upon integration.\footnote{Since $T_{i}^{++}(\alpha)$, $i\in\{0,1\}$ are superpositions of functions that are odd in $\alpha$, then an even (odd) number of derivatives with respect to $\alpha$ renders an odd (even) function with respect to that variable. Note also that $\Da$ is odd in $\alpha$, as opposed to $\Fba$.} As a consequence, the above expression simplifies to
\begin{widetext}
  \begin{align}
    \sum_{ \substack{j=0\,,\\j\textrm{ even}} }^\infty h_j^{++}\frac{1}{\xi^j} = & \int_{-1}^1d\alpha\ \left[ T_0^{++}(\alpha,t/\scale^2) + \frac{t}{\scale^2}T_1^{++}(\alpha) \right]\Da - \frac{4M^2-t}{\scale^2}\intba \beta\Fba T_1^{++\,(1)}(\alpha) \nonumber\\
                                                                                 & + \sum_{ \substack{n=2\,,\\n\textrm{ even}} }^\infty \frac{1}{\xi^n} \intba \Fba\Bigg[ \beta^{n-1} \left\{ \frac{T_0^{++\,(n-1)}(\alpha,t/\scale^2)}{(n-1)!} - \frac{t}{\scale^2}\frac{n+2}{2\cdot(n-2)!} T_1^{++\,(n-1)}(\alpha) \right\} \nonumber\\
                                                                                 & \phantom{+ \sum_{ \substack{n=2\,,\\n\textrm{ even}} }^\infty \frac{1}{\xi^n} \intba \Fba\Bigg[]} - \beta^{n+1}\frac{2M^2-t/2}{\scale^2}\frac{n+2}{n!}T_1^{++\,(n+1)}(\alpha)  \Bigg]\,,
  \end{align}
\end{widetext}
from where the first line, after restoring $t$-dependence, provides the subtraction constant of the dispersion relation:
\begin{align}
  \label{eq:h0++_tw4_full}
  h_0^{++}(t) = & \int_{-1}^1d\alpha\  T_2^{++}\left(\alpha,\frac{t}{\scale^2}\right) \Dat \nonumber\\
                & - 4\frac{M^2-t/4}{\scale^2}\intba \Fbat\beta\,T_1^{++\,(1)}(\alpha)\,,
\end{align}
where
\begin{align}
  \label{eq:DefT2}
   T_2^{++}\left(\alpha,\frac{t}{\scale^2}\right) =  T_0^{++}(\alpha,t/\scale^2) + \frac{t}{\scale^2}T_1^{++}(\alpha),
\end{align}
and the coefficients for even $n\ge 2$
\begin{align}
  \label{eq:hn++_DD}
  h^{++}_n = & \intba \Fba\Bigg[ \beta^{n-1} \left\{ \frac{T_0^{++\,(n-1)}(\alpha,t/\scale^2)}{(n-1)!}  \right . \nonumber \\
  & \left. - \frac{t}{\scale^2}\frac{n+2}{2\cdot(n-2)!} T_1^{++\,(n-1)}(\alpha) \right\} \nonumber\\
    &  - \beta^{n+1}\frac{M^2-t/4}{\scale^2}\frac{2(n+2)}{n!} T_1^{++\,(n+1)}(\alpha)  \Bigg]\,.
\end{align}
For odd $n$, $h_n^{++} = 0$\,. That only even $n$ contributes to the CFFs is a consequence of the time-reversal symmetry of the theory which leads to $\cffH^{++}(\xi) = \cffH^{++}(-\xi)$\,.

Owning to the Schwartz's reflection principle, $h^{++}_j$s must be real numbers.
This implies that only the real part of the coefficient functions should contribute to the above integrals.
In fact, the integration with respect to $\alpha$ is restricted to the interval $\alpha\in(-1,1)$.
Taking into account that $\alpha = x/\xi$, we conclude that $x$ falls in the ERBL region ($|x| < |\xi|$) while the imaginary parts of $T_0^{++}$ and $T_1^{++}$ is found in the DGLAP domain ($|x| > |\xi|$).
As a consequence, the imaginary parts of these coefficients do not contribute to the $h_j^{++}$ factors.
Note also that the argument of logarithms and dilogarithms in $T_1^{++}$ is $y=(1\pm\alpha)/2\in(0,1)$ for $\alpha\in(-1,1)$, so $\ln|y|=\ln y$ and branch cuts are not crossed for either $\ln y$ or $\Li{2}{y}$.

Equation \eqref{eq:h0++_tw4_full} deserve comments.
The kernel relating $h_0^{++}$ and $D$ is, as expected, modified by the kinematic power corrections, adding an explicit dependence in $t/\scale^2$ and a convolution with a $\mathrm{Li_2}$ function.
However, the unexpected output relies in the new mixing with the DD $F$.
This mixing is not suppressed at small values of $t$, as it comes with an explicit mass dependence.
And in fact, the prefactor $4M^2 /\scale^2$ is not small, especially for JLab kinematics, thus this term cannot be considered negligible and needs to be taken into account.
It breaks the simple\footnote{As strongly emphasised in \cite{Dutrieux:2024bgc}, if the $h_0^{++}$ is at leading twist provided by the convolution with a $D$-term, the deconvolution problem remains very challenging, with the possibility to reconstruct only a single Gegenbauer mode for now.} relation between the subtraction constant--$h_0^{++}$--and a convolution with $D$.
Worse, this term is ``unprotected'', in the sense that it is hadron-dependent, compared to standard kinematic twist expansion of the type $\xi^2 M^2/\scale^2$.
It triggers that, for ${}^4\textrm{He}$, this mass term is expected to be by far the dominant contribution, most probably precluding the extraction of the $D$-term as it was envisioned in the literature \cite{Fucini:2021psq}.
This is, provided that the dispersion relation holds for ${}^4\textrm{He}$ despite the breaking of Eq.~\eqref{eq:DRCondition}, which is only a sufficient but not strictly necessary condition.
On the other hand, the case of the pion is expected to be much better, which may allow a study of the $D$-term through the Sullivan process \cite{Amrath:2008vx,Chavez:2021koz,Chavez:2021llq,Castro:2025rpx}.


\section{Power corrections to dispersion relations: the spin-1/2 case}
\label{sec:nucleon}
In this section, we generalise the previous discussion from (pseudo-)scalar to spin-$1/2$ targets.
As the derivation in Sec.~\ref{sec:DRproof} is independent of the spin of the target, Eq.~\eqref{eq:DR_Standard} still holds for the nucleon amplitudes.
We will thus focus on the expression of $h_0^{++}$ in terms of DDs.

\subsection{Coefficient function for spin-$1/2$ targets}

Kinematic higher-twist corrections are associated to the twist decomposition of the parton operators describing the hadronic structure and, therefore, being affected by features such as spin: different spin renders different GPD parameterisation, cf.~appendix~\ref{app:spin-1/2_kernels}.
In Ref.~\cite{Braun:2025xlp}, authors present a calculation of the Compton tensor, $T^{\mu\nu}$, in DVCS for a spin-1/2 target up to kinematic twist-6.
We are interested in the transverse-helicity conserving amplitude $\amp^{++}$ from the vector part of the Compton tensor (which is a combination of the corresponding vector $\cffH^{++}, \cffE^{++}$ Compton form factors),
\begin{equation}\label{eq:ComptonTensor_sketch}
    T^{\mu\nu} = -g^{\mu\nu}_{\perp} \amp^{++} + \mathrm{(terms \sim \amp^{+-},\amp^{0+})} + (\textrm{axial part}) \,,
\end{equation}
which in Ref.~\cite{Braun:2025xlp} is given in terms of two invariant amplitudes $V_0^{(1)}$ and $V_0^{(2)}$. We would like to match that expression to the usual CFFs $\cffH^{++}$ and $\cffE^{++}$~\cite{Diehl:2003ny}, this is:
\begin{equation}
  \label{eq:ComptonTensor}
    \amp^{++} = \frac{v\cdot q'}{q\cdot q'}V_0^{(1)} + \frac{v\cdot \bp}{M^2}V_0^{(2)} = h\cffH^{++} + e\cffE^{++}\,.
\end{equation}
The amplitudes $V_0^{(1)}$ and $V_0^{(2)}$ read
\begin{widetext}
	%
		\begin{align}
			V^{(1)}_0 &=
			-\left(1+\frac{t}{4(qq')}\right)  \Big(G^{(+)}\otimes \texttt{T}_0\Big) -\frac{t}{2(qq')} \Big(G^{(+)}\otimes \texttt{T}_{10}\Big)
			 -\frac12 D_\xi^2 \frac{|\bar{p}_\perp|^2}{(qq')}
			\Big(G^{(+)}\otimes \texttt{T}_2\Big)
			+ O(\textrm{tw-6})\,,
			\\
			V^{(2)}_0 &=
			-\left(1+\frac{t}{4(qq')}\right)  \Big(\frac{E^{(+)}}{2}\odot \texttt{T}_0\Big) -\frac{t}{2(qq')}\Big(\frac{E^{(+)}}{2}\odot \texttt{T}_{10}\Big) 
			-\frac12 D_\xi \frac{|\bar{p}_\perp|^2}{(qq')}
			D_\xi \Big(\frac{E^{(+)}}{2}\odot \texttt{T}_2\Big)
			-\frac{M^2}{qq'} { D_\xi \Big(G^{(+)} \otimes \texttt{T}_2 \Big)} + O(\textrm{tw-6})\,,
		\end{align}
\end{widetext}
where $D^n_\xi = (-2\xi^2\dxi)^n$ and the hard kernels
\begin{align}\label{eq:BraunYiMansahovTw4Kernels}
	\texttt{T}_0(z) & = \frac{1}{1-z} = 2C_0(2(z-i0)-1) \,,\nonumber\\
	\texttt{T}_{10}(z) & = \frac{1}{z}\ln(1-z) = -\wpbbiii(2(z-i0)-1) \,,\nonumber\\
	\texttt{T}_2(z) & = \frac{1}{1-z}\left( \Li{2}{z} - \Li{2}{1} \right) - \frac{1}{2z}\ln(1-z) \nonumber \\
	& = \frac{\wpbbiii(2(z-i0)-1) - \mathcal{L}(2(z-i0)-1)}{2} \,.
\end{align}
Here, $z=\frac{x+\xi}{2\xi}+i0$, thus $2(z-i0)-1 = \frac{x}{\xi}$\,.
The ``magnetic'' GPD was defined\footnote{In Ref.~\cite{Braun:2025xlp} the odd in $x$ ``magnetic'' GPD was represented as $M$ so that the mapping to the notation in this manuscript is $M\to G^{(+)}$.}
as in Ref.~\cite{Braun:2025xlp} (see also Refs.~\cite{Belitsky:2005qn,Radyushkin:2013hca,Mezrag:2013mya}):
\begin{equation}
  \label{eq:magneticGPD}
    G^{(+)} = \frac{1}{2}\left( H^{(+)} + E^{(+)} \right)\,, 
\end{equation}
with $F^{(+)}(x) = F(x)-F(-x)$, and the symbols $\otimes$, $\odot$ stand for convolutions between hard coefficient kernels and GPDs with different normalisations~\cite{Braun:2025xlp}:
\begin{align}
  \label{eq:otimesDef}
  G^{(+)}\otimes \texttt{T}& = \int_{-1}^1dx\ G^{(+)}(x,\xi,t) \texttt{T}\left( \frac{x+\xi}{2\xi} + i0 \right) \,, \\
  \label{eq:odotDef}
    E^{(+)}\odot \texttt{T} & = \frac{1}{2\xi}\int_{-1}^1 dx\ E^{(+)}(x,\xi,t)\texttt{T}\left( \frac{x+\xi}{2\xi} + i0 \right)\,.
\end{align}
Meanwhile, vectors and spinor bilinears in Eq.~(\ref{eq:ComptonTensor}) are given by
\begin{align}
    v^\mu & = \bu(p')\gamma^\mu u(p) \,,  & \bp^\mu & = \frac{p+p'}{2} = \frac{\left( n'^\mu - \frac{t}{\scale^2} n^\mu\right)}{2\xi} + \bp_\perp^\mu \,, \nonumber\\
    h & = \frac{vn}{2\bp^+}\,, & e & = \frac{\bu(p') i\sigma^{\alpha\beta}n_\alpha\Delta_\beta u(p)}{4M\bp^+} \,,
\end{align}
with $\sigma^{\alpha\beta} = i\left[ \gamma^\alpha,\gamma^\beta \right]/2$. Using the definition of the skewness $\xi = -\Delta \cdot  n/(2\bp \cdot n)$ and $n=q'$, we find:
\begin{equation}
    \frac{v\cdot q'}{q\cdot q'} = \frac{v\cdot n}{-2\bp^+\xi} = -\frac{1}{\xi}h \,.
\end{equation}
With the Dirac equation
\begin{equation}
    v\cdot \bp = \bu(p')\slashed{\bp}u(p) = M\bu(p') u(p) \,,
\end{equation}
and Gordon's identity,
\begin{equation}
    v^\alpha = \bu(p')\gamma^\alpha u (p) = \bu(p')\left[ \frac{\bp^\alpha}{M} + \frac{i\sigma^{\alpha\beta}\Delta_\beta}{2M} \right] u(p) \,,
\end{equation}
we find:
\begin{align}
  vn = \frac{\scale^2}{4\xi M}\frac{v\bp}{M} + 2\bp^+ e & \Rightarrow \frac{vn}{2\bp^+} = h = \underbrace{\frac{\scale^2}{8\xi \bp^+}}_{1/2} \frac{v\bp}{M^2} + e \nonumber \\
  & \Rightarrow h-e =  \frac{v\bp}{2M^2} \,.
\end{align}
Finally,
\begin{equation}
    \amp^{++} = h\left( 2V_0^{(2)} - \frac{1}{\xi}V_0^{(1)} \right) + e\left( -2V_0^{(2)} \right) \,,
\end{equation}
where we identify
\begin{equation}
    \cffH^{++} = 2V_0^{(2)} - \frac{1}{\xi}V_0^{(1)} \,,\quad \cffE^{++} = -2V_0^{(2)} \,.
\end{equation}
$V_0^{(1)}$ is given as a convolution of some kernel with GPDs.
  As these kernels are analytic in the domain \eqref{eq:domXi}, $V_0^{(1)}$ can be expanded in powers of $1/\xi$.
  As the CFFs are even in $\xi$, it follows that $V_0^{(1)}/\xi$ can be expressed as a series in even powers of the inverse of the skewness within the domain \eqref{eq:domXi}.
  Therefore, only $V_0^{(2)}$  contributes to the subtraction constant of both $\cffH^{++}$ and $\cffE^{++}$ .
Thus,
\begin{equation}\label{eq:subtractionConstantsCancelation}
    h_0^{++} + e_0^{++} = 0 \,.
\end{equation}
Now, we want to write down the CFFs in a similar way as we did for the spin-0 target.
In order to do so we need to translate the terms with total derivatives $D^n_\xi = (-2\xi^2\dxi)^n$ to terms with $\xi^n\dxi^n$.

For that purpose, and denoting by $\texttt{T}$ the hard coefficients functions of Eqs.~(\ref{eq:BraunYiMansahovTw4Kernels}), we employ the following relations:
\begin{widetext}
  \begin{align}
    \label{eq:D2Def}
    D^2_\xi \left(\frac{|\bp_\perp|^2}{qq'} G^{(+)}\otimes \texttt{T}\right) & = 4\frac{t}{\scale^2}G^{(+)}\otimes \texttt{T} + \frac{2\xi^2\bp_\perp^2-t}{\scale^2}8\xi\dxi \left(G^{(+)}\otimes \texttt{T}\right) + \frac{8\xi^2\bp_\perp^2}{\scale^2}\xi^2\dxi^2 \left(G^{(+)}\otimes \texttt{T}\right) \,,\\
    \label{eq:DDef}
    D_\xi \left[\frac{|\bp_\perp|^2}{qq'}D_\xi\left( \frac{E^{(+)}}{2}\odot \texttt{T} \right) \right] & = \frac{2\xi^2\bp_\perp^2-t}{\scale^2}8\xi\dxi \left( \frac{E^{(+)}}{2}\odot \texttt{T} \right) + \frac{8\xi^2\bp_\perp^2}{\scale^2} \xi^2\dxi^2 \left( \frac{E^{(+)}}{2}\odot \texttt{T} \right) \,,
  \end{align}
\end{widetext}
Note also that in Ref.~\cite{Braun:2025xlp}, the notation $H,E,\widetilde{H},\widetilde{E}$ refers in fact to half of the C-even part of the GPDs as introduced in~\cite{Diehl:2003ny}, this is $H,E,\widetilde{H},\widetilde{E}\mapsto H^{(+)}/2,E^{(+)}/2,\widetilde{H}^{(+)}/2,\widetilde{E}^{(+)}/2$.

Collecting the above results,
\begin{widetext}
	\begin{align}
		\cffH^{++} = &\ -\left( 1 - \frac{t}{2\scale^2} \right)\left[ E^{(+)}\odot \texttt{T}_0 - \frac{1}{\xi}G^{(+)}\otimes \texttt{T}_0 \right] +\frac{t}{\scale^2}\left[ E^{(+)}\odot \texttt{T}_{10} - \frac{1}{\xi} G^{(+)}\otimes \texttt{T}_{10} \right] \nonumber\\
		&\ - \frac{1}{2} \Bigg[ \frac{2\xi^2\bp_\perp^2-t}{\scale^2}8\xi \left(\dxi \left[E^{(+)}\odot \texttt{T}_2\right] - \frac{1}{\xi}\dxi \left[G^{(+)}\otimes \texttt{T}_2\right] \right) + \frac{8\xi^2\bp_\perp^2}{\scale^2}\xi^2 \left(\dxi^2 \left[E^{(+)}\odot \texttt{T}_2\right] - \frac{1}{\xi}\dxi^2 \left[G^{(+)}\otimes \texttt{T}_2\right] \right) \Bigg] \nonumber\\
		&\ + \frac{1}{\xi}\frac{2t}{\scale^2}G^{(+)}\otimes \texttt{T}_2 - \frac{8M^2}{\scale^2}\xi^2\dxi \left(G^{(+)}\otimes \texttt{T}_2\right) \,.
	\end{align}
\end{widetext}
Making use of the previously introduced expressions for $\texttt{T}_0$, $\texttt{T}_{10}$ and $\texttt{T}_2$, we find the following relations to the convolutions encountered in the spin-0 case. Starting with the term free of derivatives:
\begin{align}
  & E^{(+)}\odot \texttt{T}_0 - \frac{1}{\xi}G^{(+)}\otimes \texttt{T}_0 \nonumber \\
  = & -\int_{-1}^1dx\ \frac{1}{\xi} C_0(x/\xi)H^{(+)}(x,\xi,t) \,,
\end{align}
and
\begin{align}
  & E^{(+)}\odot \texttt{T}_{10} - \frac{1}{\xi}G^{(+)}\otimes \texttt{T}_{10}  \nonumber \\
  = & \int_{-1}^1 dx\ \frac{1}{2\xi}\wpbbiii(x/\xi) H^{(+)}(x,\xi,t) \,.
\end{align}
The first derivative term yields:
\begin{align}
& \dxi \left(E^{(+)} \odot \texttt{T}_2\right) - \frac{1}{\xi}\dxi \left( G^{(+)}\otimes \texttt{T}_2 \right) \nonumber \\
  = & -\frac{1}{2}\left( \frac{1}{\xi}E^{(+)}\otimes \texttt{T}_2 + \dxi \left[H^{(+)}\otimes \texttt{T}_2\right] \right) \nonumber\\
  = &-\frac{1}{4}\int_{-1}^1 \frac{\textrm{d}x}{\xi}\Bigg(\frac{1}{\xi}\left[ \wpbbiii(x/\xi) - \mathcal{L}(x/\xi) \right]E^{(+)}(x,\xi,t) \nonumber\\
    & \phantom{= -\frac{1}{4}\int_{-1}^1 } + \dxi \left\{ \left[ \wpbbiii(x/\xi) - \mathcal{L}(x/\xi) \right]H^{(+)}(x,\xi,t) \right\} \Bigg) \,,\nonumber\\
\end{align}
while the second derivative term gives:
\begin{align}
  &  \dxi^2 \left( E^{(+)}\odot \texttt{T}_2 \right) - \frac{1}{\xi}\dxi^2 \left( G^{(+)}\otimes \texttt{T}_2 \right) \nonumber \\
  = & \frac{1}{\xi^3}\left( E^{(+)}\otimes \texttt{T}_2 - \xi\dxi \left[E^{(+)}\otimes \texttt{T}_2\right] - \frac{\xi^2}{2}\dxi^2 \left[H^{(+)}\otimes \texttt{T}_2\right] \right) \nonumber \\
   = &\frac{1}{2\xi^2}\int_{-1}^1\frac{\textrm{d}x}{\xi}\Bigg( \left[ \wpbbiii(x/\xi) - \mathcal{L}(x/\xi) \right]E^{(+)}(x,\xi,t) \nonumber\\
    & \phantom{= \frac{1}{2\xi^2}} -\xi\dxi\left\{ \left[ \wpbbiii(x/\xi) - \mathcal{L}(x/\xi) \right]E^{(+)}(x,\xi,t) \right\} \nonumber\\
    & \phantom{= \frac{1}{2\xi^2}} -\frac{\xi^2}{2}\dxi^2 \left\{ \left[ \wpbbiii(x/\xi) - \mathcal{L}(x/\xi) \right]H^{(+)}(x,\xi,t) \right\} \Bigg)\,.
\end{align}
As a consequence, the CFF $\cffH^{++}$ for the spin-1/2 case can be written in the compact form
\begin{equation}
  \label{eq:cffH_spin1/2_decomposition}
    \cffH^{++} = \mathbb{F}_0[H^{(+)}] + \Delta\cffH^{++} + O(\textrm{tw-6})\,,
\end{equation}
where $\mathbb{F}_0[H^{(+)}]$ is the convolution of the spin-0 case~\eqref{eq:cffH_full_compact} and 
\begin{align}
  \label{eq:Delta_cffH}
    \Delta\cffH^{++}  = & \int_{-1}^1\frac{\textrm{d}x}{\xi} \left(\frac{-t}{2\scale^2}\right) \Bigg\{ \left( \wpbbiii - \mathcal{L} \right)E^{(+)} \nonumber\\
    & \phantom{\int_{-1}^1dx} - \xi\dxi\left[\left( \wpbbiii - \mathcal{L} \right)E^{(+)}\right] \nonumber\\
    & \phantom{\int_{-1}^1dx} + \textcolor{blue}{\xi^3}\dxi \left[\left( \wpbbiii - \mathcal{L} \right)\left(H^{(+)} + E^{(+)}\right) \right] \Bigg\} \,.
\end{align}
Following the same procedure and after some algebra,
\begin{equation}
    \cffE^{++} = \mathbb{F}_0[E^{(+)}] + \Delta\cffE^{++} + O(\textrm{tw-6}) \,,
\end{equation}
where $\mathbb{F}_0[E^{(+)}]$ corresponds to the convolution introduced in Eq.~\eqref{eq:cffH_full_compact} but changing the GPD $H^{(+)}$ by $E^{(+)}$, and $\Delta\cffE^{++}$ is
\begin{align}
  \Delta\cffE^{++}  = &\int_{-1}^1dx\ \frac{1}{\xi}\left(\frac{t}{2\scale^2}\right) \Bigg\{ \left( \wpbbiii - \mathcal{L} \right)E^{(+)} \nonumber \\
  & \quad - \xi\dxi\left[ \left( \wpbbiii - \mathcal{L} \right)E^{(+)} \right] \nonumber\\
                      & \quad  + 2 \xi^3\dxi\left[ \left( \wpbbiii - \mathcal{L} \right)G^{(+)}\right] \nonumber \\
  & \quad - 2 \xi\dxi\left[ \left( \wpbbiii - \mathcal{L} \right)G^{(+)} \right] \Bigg\} \nonumber\\
    & + \int_{-1}^1\frac{\textrm{d}x}{\xi} \frac{4\xi^3\bp^2_\perp}{\scale^2}\dxi\left[ \left(\wpbbiii-\mathcal{L}\right)G^{(+)} \right] \,.
\end{align}
Note that beyond LT there is no 1-to-1 relation between GPDs $H,E$ and CFFs $\cffH^{++},\cffE^{++}$. However, considering the combination given by the ``magnetic'' GPD~\eqref{eq:magneticGPD} a 1-to-1 relation still holds:
\begin{align}
  \label{eq:magneticCFF}
    \cffG^{++} & = \frac{1}{2}\left( \cffH^{++}+\cffE^{++} \right) \nonumber\\
    & = \mathbb{F}_0[G^{(+)}] - 4\int_{-1}^1 \frac{\textrm{d}x}{\xi} \frac{M^2-t/4}{\scale^2}\xi^3\dxi\left( \frac{\mathcal{L}-\wpbbiii}{2} G^{(+)} \right) \,.
\end{align}

\subsection{Subtraction constant at twist-4 for spin-1/2 targets}

Taking into account that the CFF $\cffH^{++}$ of a spin-1/2 hadron can be decomposed into the convolution of a spin-0 particle, $\mathbb{F}_0[\cffH^{(+)}]$~(\ref{eq:cffH_full_compact}), and an addendum $\Delta\cffH^{++}$~(\ref{eq:Delta_cffH}), we can profit from the previous calculation and focus on the latter term. In the notation of the preceding section, 
\begin{equation}
    \Delta\cffH^{++} = \sum_{i=0}^2 \Delta\cffH^{++}_i\,,
\end{equation}
where
\begin{align}
  \label{eq:Delta_cffH_0}
    \Delta\cffH^{++}_0  = &\frac{t}{\scale^2}\int_{-1}^1 dx\ \frac{1}{\xi} T_1^{++}E \nonumber\\
  = &\frac{t}{\scale^2}\left[ \sum_{n=0}^\infty \frac{1}{n!}\intba \frac{\beta^n}{\xi^{n+1}} T_1^{++\,(n)}(\alpha)\Kbat \right. \nonumber \\
  & - \left. \int_{-1}^1d\alpha\ T_1^{++}(\alpha)D(\alpha) \right] \,,
\end{align}
\begin{align}
  \label{eq:Delta_cffH_1}
  \Delta\cffH^{++}_1  = & -\frac{t}{\scale^2} \int_{-1}^1 dx\ \dxi\left( T_1^{++}E \right) \nonumber\\
  = &\frac{t}{\scale^2}\left[ \sum_{n=0}^\infty \frac{1}{n!}\intba \frac{\beta^{n+1}}{\xi^{n+2}} T_1^{++\,(n+1)}(\alpha) \Kbat\right. \nonumber \\
  & \left. + \int_{-1}^1 d\alpha\ T_1^{++}(\alpha)D(\alpha) \right] \,,
\end{align}
and
\begin{align}
  \label{eq:Delta_cffH_2}
    \Delta\cffH^{++}_2 & = \int_{-1}^1dx\ \xi^2\dxi\left( 2 T_1^{++} G \right)  \nonumber\\
    & = \frac{-t}{\scale^2} \sum_{n=0}^\infty \frac{1}{n!}\intba \frac{\beta^{n+1}}{\xi^{n}} 2 T_1^{++\,(n+1)}(\alpha) \Nbat \,.
\end{align}
The zeroth order in the expansion of $\Delta\cffH^{++}(\xi,t)$ in the analytical unphysical domain of the skewness is then
\begin{align}
  \left.\Delta\cffH^{++}\right|_{1/\xi^0} = -\frac{t}{\scale^2}\intba&  \beta T_1^{++\,(1)}(\alpha) \nonumber \\
  &  \times \left[ \Fbat+\Kbat \right]\, .
\end{align}
Adding this result to the corresponding term $\left.\mathbb{F}_0[H^{(+)}]\right|_{1/\xi^0}$ which takes the form of Eq.~\eqref{eq:h0++_tw4_full}, we find the subtraction constant for the dispersion relation associated to the spin-1/2 hadron in DVCS:
\begin{widetext}
  \begin{align}
    \label{eq:h0++_tw4_full_spin1/2}
        h_0^{++}(t) = & \int_{-1}^1d\alpha\ \left[ T_0^{++}(\alpha,t/\scale^2) + \frac{t}{\scale^2}T_1^{++}(\alpha) \right] \Dat \nonumber\\
        & - 4\frac{M^2}{\scale^2}\intba \left[\Fbat + \frac{t}{4M^2}\Kbat \right] \beta\,T_1^{++\,(1)}(\alpha)\,.
    \end{align}
\end{widetext}
Taking $K\mapsto -F$ recovers the subtraction constant of the spin-0 case.

We could follow the same steps for the CFF $\cffE^{++}$ and obtain
\begin{equation}
    e_0^{++} = -h_0^{++}\,,
\end{equation}
which holds up to twist-6 at least, as discussed in Eq.~\eqref{eq:subtractionConstantsCancelation}. This has been cross-checked by computing the corresponding $g_0^{++} = (h_0^{++}+e_0^{++})/2 = 0$ from Eq.~\eqref{eq:magneticCFF}.

The remnant coefficients for the series in powers of $1/\xi$ of the CFF $\cffH^{++}$ are, for even $n \geq 2$,
\begin{widetext}
  \begin{align}
    \label{eq:hn++_tw4_full_spin1/2}
    h_n^{++} & = \intba \Fbat \Bigg[ \beta^{n-1}\left\{ \frac{T_0^{++\,(n-1)}(\alpha,t/\scale^2)}{(n-1)!} - \frac{t}{\scale^2}\frac{n+2}{2\cdot(n-2)!}T_1^{++\,(n-1)}(\alpha) \right\} \nonumber\\
             & \phantom{= \intba \Fbat \Bigg[\ } -\beta^{n+1}\left\{ \frac{M^2-t/4}{\scale^2} \frac{1}{(n-1)!} + \frac{M^2}{\scale^2}\frac{2}{n!} \right\} 2T_1^{++\,(n+1)}(\alpha) \Bigg] \nonumber\\
             & \phantom{=\ } + \intba \Kbat \frac{t}{\scale^2} \left[ \beta^{n-1}\frac{n}{(n-1)!}T_1^{++\,(n-1)}(\alpha) - \beta^{n+1}\frac{1}{n!}T_1^{++\,(n+1)}(\alpha) \right] \,,
  \end{align}
\end{widetext}
and for odd $n$, $h^{++}_n=0$\,.

These coefficients together with those for the magnetic CFF can be used to compute those of $\cffE^{++}$. For $\cffG^{++}$ and even $n\geq 2$,
\begin{align}
  g_n^{++}  = &\intba \Nbat\, 2 \Bigg[ \beta^{n-1}\left\{ \frac{T_0^{++\,(n-1)}(\alpha,t/\scale^2)}{(n-1)!} \right.\nonumber \\
  & \phantom{ \intba }- \left. \frac{t}{\scale^2}\frac{n+2}{2\cdot(n-2)!}T_1^{++\,(n-1)} \right\} \nonumber\\
    & \phantom{ \intba } -\beta^{n+1}\frac{M^2-t/4}{\scale^2}\frac{2n}{n!}T_1^{++\,(n+1)}(\alpha) \Bigg] \,,
\end{align}
and for odd $n$, $g_n^{++}=0$ as usual.

The dispersion relation connecting $D(\alpha)$ to $h_0^{++}$ suffers the same issue than in the scalar case, with an additional complication: it also mixes GPDs $H$ and $E$ through DDs $F$ and $K$.
Indeed, on top of GPD $H$ as in the scalar case, GPD $E$ also contributes to the dispersion relation.
This triggers a new challenge, as $E$ is poorly known today, adding uncertainties on the extraction of $D$.

\subsection{Comparison with BMMP results}

\begin{figure}[t]
  \centering
  \includegraphics[width=0.5\textwidth]{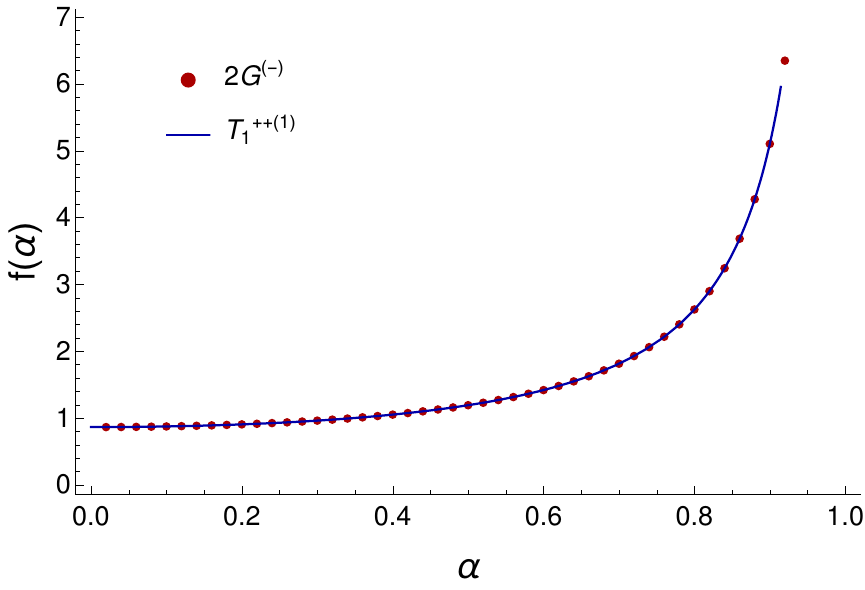}
  \caption{Comparison between the present result $T_1^{++(1)}$ and the first result $2G^{(-)}$ obtained in Ref.~\cite{Braun:2014sta}.}
  \label{fig:BMPvsMM}
\end{figure}

In this section, we compare the results of Eq.~\eqref{eq:h0++_tw4_full_spin1/2} with the pioneering one obtained in Ref.~\cite{Braun:2014sta}.
First, let us recall the expression for the subtraction constant obtained in \cite{Braun:2014sta}:
\begin{align}
  \label{eq:BMMPResult}
  h_0^{++} = &  2 \int_0^1 \frac{\textrm{d}u}{1-u}\left(1 - \frac{t}{Q^2}\left(1-2\ln u \right) \right)\varphi_D(u) \nonumber \\
             & -4\int_0^1d\xi\ \xi\int_\xi^1 \frac{dx}{x}\ t_2(x) \nonumber \\
  & \quad \times \left[ \frac{4M^2}{Q^2}H^{(+)}(\xi/x,\xi,t) + \frac{t}{Q^2}E^{(+)}(\xi/x,\xi,t)  \right]\,,
\end{align}
where
\begin{align}
  \label{eq:t2Def}
  t_2(x) = \frac{\ln \left(\frac{1+x}{2x} \right)}{1-x}-\frac{1}{2(1+x)}\, ,
\end{align}
and
\begin{align}
  \varphi_D(u) = D(2u-1)\, .  
\end{align}
The first line of Eq.~\eqref{eq:BMMPResult} can be related to our result by making the change of variable $u\to \frac{1+\alpha}{2}$ and antisymmetrizing the coefficient function.
The second line is more challenging.
Since the structure in terms of powers of $M^2/Q^2$ and $t/Q^2$ is the same than in our case\footnote{We recall the reader that the difference between choosing $Q^2$ or $\scale^2$ is a twist-6 effect.} we will only focus on $H^{(+)}$.
Since
\begin{align}
  \label{eq:DefFplus}
  H^{(+)}(x,\xi,t) = & \intba \deltaxba\nonumber\\
  & \quad\times \left(F(\beta,\alpha,t)-F(-\beta,\alpha,t)\right)\nonumber \\
  = & \intba \deltaxba F^{(+)}(\beta,\alpha,t) \, ,
\end{align}
we can inject this DD representation into the $H^{(+)}$ integral of~\eqref{eq:BMMPResult}:
\begin{align}
  I_2 = &  4\int_0^1d\xi\ \xi\int_\xi^1 \frac{dx}{x}\ t_2(x) H^{(+)}(\xi/x,\xi,t) \nonumber \\
  = & 4 \intba F^{(+)}(\beta,\alpha,t)  \nonumber \\
  & \quad \times \int_0^1\textrm{d}\xi\ \xi\int_\xi^1 \frac{\textrm{d}x}{x}\ t_2(x) \delta\left(\frac{\xi}{x}-\beta-\alpha\xi\right).
\end{align}
We then perform the change of variable $y = \xi / x$ and integrate the Dirac Delta over $y$:
\begin{align}
  \label{eq:I2CoV}
  I_2 = & 4 \intba F^{(+)}(\beta,\alpha) \int_0^1\textrm{d}\xi\ \frac{\xi}{\beta + \alpha \xi}  t_2\left(\frac{\xi}{\beta+\alpha \xi}\right) \nonumber \\
  & \quad \times \Theta\left(\xi\le \beta+\xi\alpha \le 1 \right)
\end{align}
where $\Theta$ is non-zero only for $\xi\le \beta+\xi\alpha \le 1$.
Because of the DD support $\Omega$, the upper bound does not provide additional constraint.
Since $\xi \ge 0$, the lower bound can be understood as $0 \le \xi \le \beta /(1-\alpha)$ and $\beta \ge 0$.
Re-injecting this condition in Eq.~\eqref{eq:I2CoV}, we obtain:
\begin{align}
  \label{eq:I2Theta}
  I_2 = & 4 \intba F^{(+)}(\beta,\alpha) \Theta(\beta) \int_0^{\frac{\beta}{1-\alpha}}\hspace{-2mm}  \frac{\textrm{d}\xi~\xi}{\beta + \alpha \xi}  t_2\left(\frac{\xi}{\beta+\alpha \xi}\right).
\end{align}
Now performing the change of variable $\xi = \beta z$, we are left with:
\begin{align}
  \label{eq:I2GoodBeta}
  I_2 = & 4 \intba F^{(+)}(\beta,\alpha) \Theta(\beta)\beta \int_0^{\frac{1}{1-\alpha}} \hspace{-2mm}\frac{ \textrm{d}z~z}{1 + \alpha z}  t_2\left(\frac{z}{1+\alpha z}\right)\nonumber \\
     = &  \intba \beta F(\beta,\alpha)   4 \int_0^{\frac{1}{1-\alpha}}\textrm{d}z  \frac{z}{1 + \alpha z}  t_2\left(\frac{z}{1+\alpha z}\right),
\end{align}
where in the second line we have exploited the definition of $F^{(+)}$ to remove the $\Theta$ function and come back to $F$.
Introducing now $G(\alpha)$:
\begin{align}
  \label{eq:DefG}
  G(\alpha) =  \int_0^{\frac{1}{1-\alpha}}\textrm{d}z  \frac{z}{1 + \alpha z}  t_2\left(\frac{z}{1+\alpha z}\right),
\end{align}
and symmetrising it,
\begin{align}
   G^{(-)}(\alpha) =  G(\alpha) +  G(-\alpha), 
\end{align}
we are left with
\begin{align}
  \label{eq:FinalI2}
    I_2 =  \intba \beta F(\beta,\alpha) 2  G^{(-)}(\alpha) ,
\end{align}
an analytic structure in agreement with our result in Eq.~\eqref{eq:h0++_tw4_full_spin1/2}.
The remaining task is to prove that:
\begin{align}
  \label{eq:LastProof}
  2  G^{(-)}(\alpha) = T_1^{++\,(1)}(\alpha).
\end{align}
We provide here a numerical comparison on Fig.~\ref{fig:BMPvsMM}.
This equality is verified with a maximal relative error of $10^{-13}$ reached when $\alpha \to 1$, \emph{i.e.} when approaching the divergence.
We thus conclude that our new derivation is in agreement with the one obtained previously in Ref.~\cite{Braun:2014sta}.

Finally, let us mention that, our sufficient kinematic condition $Q^2 \ge 2M^2 - t$, is not mentioned in the proof of Ref.~\cite{Braun:2014sta}.
  Note that in our case, this condition is only sufficient,  not necessary, to guarantee that on a segment of the real axis, the amplitude is real.
  Relaxing it would requires to analyse the possible production threshold in the Mandelstam plane.


\section{Impact on the deconvolution problem}
\label{sec:Deconvolution}
In this section, we assume that $h_0^{++}$ is known experimentally and that $F$ and $K$ have already been extracted.
The question we ask is whether the modifications of the coefficient function are sufficient to allow one to deconvolute two (or more) Gegenbauer modes of the $D$-term.

For both the spin-0 and 1/2 cases, the integral containing the $D$-term is
\begin{align}
  \label{eq:Dscr}
  \mathscr{D}(t) = &  \int_{-1}^1d\alpha\ T_2^{++}(\alpha,t/\scale^2) \Dat \nonumber \\ 
   = &  \int_{-1}^1d\alpha\ \left[ T_0^{++}(\alpha,t/\scale^2) + \frac{t}{\scale^2}T_1^{++}(\alpha) \right]\Dat \,.
\end{align}
With Eqs.~\eqref{eq:C0}, \eqref{eq:wpbbiii}, \eqref{eq:calL} and \eqref{eq:T0++_and_T1++}, we can express the above term in the square brackets as
{\allowdisplaybreaks
\begin{align}
    T_2^{++}(\alpha,t/\scale^2) & = C_{\rm LT}^{(+)}(\alpha) + \frac{t}{2\scale^2}\left[ \wpbbiii^{(+)}(\alpha) - C_0^{(+)}(\alpha) \right] \label{eq:T2++} \\
    & \overset{\rm LO}{=} \left( 1-\frac{t}{2\scale^2} \right)C_0^{(+)}(\alpha) + \frac{t}{2\scale^2}\wpbbiii^{(+)}(\alpha) \,. \label{eq:T2++_LO}
\end{align}
}
To solve the integral~\eqref{eq:Dscr} with the LO kernel~\eqref{eq:T2++_LO}, we choose the traditional Gegenbauer parameterization of the $D$-term that for quarks reads
\begin{equation}
  \label{eq:GegenbauerDecomposition}
    D(\alpha,t) = (1-\alpha^2)\sum_{n=0}^\infty d_{2n+1}(t) \gegen{2n+1}{3/2}(\alpha) \,.
\end{equation}
Here, $\gegen{2n+1}{3/2}(\alpha)$ is a Gegenbauer polynomial of degree $2n+1$ in $\alpha$ .

\subsubsection{First integral}

The first part of the convolution involves:
\begin{align}
  \label{eq:firstintDterm}
  J = &\left( 1-\frac{t}{2\scale^2} \right) \nonumber \\
   & \times \sum_{n=0}^\infty d_{2n+1}(t) \int_{-1}^1 \textrm{d}\alpha C_0^{(+)}(\alpha) (1-\alpha^2) \gegen{2n+1}{3/2}(\alpha)\, ,
\end{align}
which is the same integral than in the pure LO-LT case. Thus the result is already well known and given as:
\begin{equation}
  \label{eq:firstintDtermInt}
    J = 4 \left( 1-\frac{t}{2\scale^2} \right) \sum_{n=0}^\infty d_{2n+1}(t)\, .
 \end{equation}

\subsubsection{Second integral}

The second part of Eq.~\eqref{eq:Dscr} is given by:
{\allowdisplaybreaks
  \begin{align}
    &  \int_{-1}^1 d\alpha\ \wpbbiii^{(+)}(\alpha) D(\alpha,t) \nonumber \\
    = &\int_{-1}^1d\alpha\ \left[ -\frac{2}{1+\alpha}\Ln{\frac{1-\alpha}{2}} + \frac{2}{1-\alpha}\Ln{\frac{1+\alpha}{2}} \right]D(\alpha,t) \nonumber\\
    = & \int_{-1}^1d\alpha\ \frac{4}{1-\alpha}\Ln{{\frac{1+\alpha}{2}}} D(\alpha,t) \nonumber\\
    = & 4\sum_{n=0}^\infty d_{2n+1}(t) \int_{-1}^1d\alpha\ (1+\alpha)\Ln{\frac{1+\alpha}{2}}\gegen{2n+1}{3/2}(\alpha) \,.
  \end{align}
}
The integral with respect to $\alpha$ on the RHS, 
\begin{equation}
    I_N = \int_{-1}^1d\alpha\ (1+\alpha)\Ln{\frac{1+\alpha}{2}}\gegen{N}{3/2}(\alpha) \,,
\end{equation}
can be solved for any degree $N$ of the Gegenbauer polynomial by making use of property~(\ref{prop::generating_functional}):
\begin{widetext}
  \begin{align}
    \label{eq:sum_t^NI_N}
    \sum_{N=0}^\infty \tau^N I_N = &\int_{-1}^1d\alpha\ (1+\alpha) \Ln{\frac{1+\alpha}{2}}\frac{1}{(1-2\tau\alpha+\tau^2)^{3/2}} \nonumber\\
    = & \frac{1}{\tau^2}\Bigg[ -\sqrt{1-2\tau\alpha+\tau^2} + \frac{1+\tau+\tau^2-\tau\alpha}{\sqrt{1-2\tau\alpha+\tau^2}}\Ln{\frac{1+\alpha}{2}}  \nonumber\\
     & \phantom{ \frac{1}{\tau^2}\Bigg[ } + (1+\tau)\left\{ \Ln{1+\frac{\sqrt{1-2\tau\alpha+\tau^2}}{1+\tau}} - \Ln{1-\frac{\sqrt{1-2\tau\alpha+\tau^2}}{1+\tau}} \right\} \Bigg] \Bigg|_{\alpha\to -1}^{\alpha \to +1} \nonumber\\
     = &\frac{1}{\tau^2}\Bigg[ 2\tau + (1+\tau) \Bigg\{ \ln 2 - \ln 0^+ + \Ln{\frac{2}{1+\tau}} - \ln 2 - \Ln{\frac{2\tau}{1+\tau}} + \mathbf{L} \Bigg\} \Bigg] \,,
  \end{align}
\end{widetext}
where $\ln 0^+ = \lim_{\eps\to 0^+}\ln \eps$ and
\begin{align}
    \mathbf{L} & = \lim_{\alpha\to -1^+}\Ln{1-\frac{\sqrt{1-2\tau\alpha+\tau^2}}{1+\tau}} \,.
\end{align}
Introducing $\alpha=-1+\eps,\,\eps>0$, then
{\allowdisplaybreaks
  \begin{align}
    \mathbf{L} & = \lim_{\eps\to 0^+} \Ln{\frac{1+\tau - \sqrt{1+2\tau+\tau^2-2\tau\eps}}{1+\tau}} \nonumber\\
               & = -\ln(1+\tau) + \lim_{\eps\to 0^+} \Ln{1+\tau-\sqrt{(1+\tau)^2-2\tau\eps}} \nonumber\\
               & = -\ln(1+\tau) + \lim_{\eps\to 0^+} \Ln{1+\tau-(1+\tau)\sqrt{1-\frac{2\tau\eps}{(1+\tau)^2}}} \,. 
  \end{align}
}
The function $2\tau/(1+\tau)^2$ is monotonic increasing for $\tau\in[0,1]$ taking values in $[0,1/2]$ with the maximum located at $\tau=1$. Then, we can safely consider $2\tau\eps/(1+\tau)^2 < 1$ and expand the above square root in Taylor series:
\begin{align}
  \label{eq:DefBoldL}
    \mathbf{L} & = -\ln(1+\tau) + \lim_{\eps\to 0^+} \Ln{\frac{\tau \eps}{1+\tau}} \nonumber\\
    & = -2\ln(1+\tau)+\ln\tau+\ln 0^+ \,.
\end{align}

Going back to Eq.~\eqref{eq:sum_t^NI_N},
{\allowdisplaybreaks
  \begin{align}
    \sum_{N=0}^\infty \tau^N I_N & = \frac{2}{\tau^2} \left[ \tau - (1+\tau)\ln(1+\tau) \right] \nonumber\\
                                 & = \frac{2}{\tau^2}\left[ \tau - (1+\tau)\sum_{n=0}^\infty(-1)^n\frac{\tau^{n+1}}{n+1} \right] \nonumber\\
                                 & = \frac{2}{\tau^2}\left[ \tau - (1+\tau)\left(\tau + \sum_{n=1}^\infty(-1)^n\frac{\tau^{n+1}}{n+1}\right) \right] \nonumber\\
                                 & = -2\left[ 1 - \sum_{n=0}^\infty (-1)^n \frac{\tau^n}{n+2} - \sum_{n=0}^\infty (-1)^n\frac{\tau^{n+1}}{n+2}\right] \nonumber\\
                                 & = -1 - 2\sum_{n=1}^\infty \tau^n\frac{(-1)^n}{(n+1)(n+2)} \nonumber\\
                                 & = \sum_{n=0}^\infty \tau^n\frac{(-1)^{n+1} 2}{(n+2)(n+1)}\,.
  \end{align}
}
As the equality holds for all $|\tau|\leq 1$, we conclude
\begin{align}
  I_N & = \int_{-1}^1d\alpha\ (1+\alpha)\Ln{\frac{1+\alpha}{2}}\gegen{N}{3/2}(\alpha) \nonumber \\
  & = \frac{(-1)^{N+1} 2}{(N+2)(N+1)} \,,
\end{align}
from where it follows
\begin{equation}
  \label{eq:Dscr_wpbbiii_final}
    \int_{-1}^1 d\alpha\ \wpbbiii^{(+)}(\alpha) D(\alpha,t) = \sum_{n=0}^\infty d_{2n+1}(t)\frac{4}{(2n+3)(n+1)} \,.
\end{equation}

\subsubsection*{Full integral with the $D$-term}

With the results from Eqs.~\eqref{eq:firstintDtermInt} and \eqref{eq:Dscr_wpbbiii_final}, we can finally write
\begin{align}
  \label{eq:Dscr_final}
  \mathscr{D}(t) & \overset{\rm LO}{=} 4\sum_{n=0}^\infty d_{2n+1}(t) - \frac{2t}{\scale^2}\sum_{n=0}^\infty d_{2n+1}(t) \frac{(2+n)(1+2n)}{(1+n)(3+2n)} \nonumber \\
  & \overset{\rm LO}{=}  4\sum_{n=0}^\infty d_{2n+1}(t) \left[1-\frac{t}{2\scale^2}\left(1-\frac{1}{(2n+3)(n+1)} \right) \right] .
\end{align}
Contrary to the pure LT case, there is a $n$-dependence introduced in the description of the subtraction constant in terms of Gegenbauer modes of order $n$.
However, these coefficient converge quadratically to $1$, and thus we cannot expect to distinguish the behaviour beyond the very first modes.
The $t/\scale^2$-dependence becomes degenerate.

\subsubsection{Shadow contributions to the $D$-term}

At LO in $\alpha_s$ we have found for DVCS that the integral containing the $D$-term can be analytically computed by a Gegenbauer parameterization and it takes the form of Eq.~\eqref{eq:Dscr_final}. Taking into account that $d_{2n+1}(t)$ is a shorthand for $d_{2n+1}(t;\,\mu^2)$ with $\mu^2$ the energy scale, then a term that produces a vanishing $\mathscr{D}(t;\,\mu^2)$ at a certain energy scale $\mu^2$ and a certain ratio $t/\scale^2$ is referred to as a {\it shadow $D$-term} \cite{Dutrieux:2024bgc}.

Assuming dominance by the first two Gegenbauer modes ($d_{n}(t) = 0$, $\forall n > 3$), and omitting the dependence on $\mu^2$, a shadow $D$-term is given by
\begin{equation}
    \mathscr{D}^{\rm sh}(t) = 0 = d_1^{\rm sh}(t)\left[ 4 - \frac{4}{3}\frac{t}{\scale^2} \right] + d_3^{\rm sh}(t)\left[ 4 - \frac{9}{5}\frac{t}{\scale^2} \right]\,.
\end{equation}
At LO and LT, a shadow $D$-term is manifest through the condition $d_1^{\rm sh}(t) = -d_3^{\rm sh}(t)$. For a non-zero but fixed ratio $|t|/\scale^2 < 1$, we find:
\begin{equation}
  \label{eq:shadowDterm_tw4}
    d_1^{\rm sh}(t) = -d_3^{\rm sh}(t)\frac{1-\frac{9}{20}\frac{t}{\scale^2}}{1-\frac{1}{3}\frac{t}{\scale^2}}\,.
\end{equation}
 In order to determine the impact of the kinematic higher-twist corrections on the deconvolution problem, we consider the difference:
 \begin{align}
     \frac{1-\frac{9}{20}\frac{t}{\scale^2}}{1-\frac{1}{3}\frac{t}{\scale^2}} - 1 & = \left(1-\frac{9}{20}\frac{t}{\scale^2}\right)\left(1+\frac{1}{3}\frac{t}{\scale^2} + O(|t|^2/\scale^4)\right) - 1 \nonumber\\
     & = -\frac{7}{60}\frac{t}{\scale^2} + O(|t|^2/\scale^4) \nonumber\\
     & \simeq 0.12\left| \frac{t}{\scale^2} \right| + O(|t|^2/\scale^4)\,.
 \end{align}
 This difference represents the modification on the LO+LT shadow $D$-term ($d_1^{\rm sh}(t) = -d_3^{\rm sh}(t)$) due to the kinematic power corrections. We find that said modification is of order $\sim$10\% of a twist-4, rendering the effect of the  $t/\scale^2$-corrections on the deconvolution problem probably not better than the one obtained through evolution \cite{Dutrieux:2024bgc}.
 An improvement on the extraction of the $D$-term by including these effects should not be expected, at least if the extraction is performed from the dispersion relation associated to $\cffH^{++}$ as in Ref.~\cite{Dutrieux:2024bgc}.
 This motivates the study of the dispersion relation of the other CFFs ($\cffH^{+-}$, $\cffH^{0+}$) as for those ones there is no LT component that could obscure the kinematic corrections. From Eq.~\eqref{eq:shadowDterm_tw4}, we deduce that their effect on purely higher-twist components should be of the order of 
\begin{equation}
    \frac{9/20}{1/3} = 27/20 = 1.35 \Rightarrow d_1^{\rm sh}(t) \sim 1.35 d_3^{\rm sh}(t)\,.
\end{equation}
This is an estimated 35\% difference between the first two Gegenbauer modes allowing us to study the deconvolution problem and the relation between the different modes. The computation of the dispersion relations associated to the CFFs $\cffH^{+-}$ and $\cffH^{0+}$ and the subsequent extraction of $D$-term will be taken care of in a next publication.


\section{Numerical evaluation}
\label{sec:Numerical}
In this section, we will assess the impact of kinematic higher-twist corrections on the description of the subtraction constant of DVCS.
To do so, we will combine the experimental extraction of $d_1$ (see Eq.~\eqref{eq:GegenbauerDecomposition}) obtained in Ref.~\cite{Dutrieux:2024bgc}, with the Goloskokov-Kroll (GK) GPD model \cite{Goloskokov:2006hr,Goloskokov:2007nt,Goloskokov:2008ib,Kroll:2012sm}.
Note that this choice is not fully consistent, in the sense that the GK model has been extracted assuming explicitly that the $D$-term vanishes.
Moreover, since the extractions performed both for $d_1$ and the GK model were performed assuming only leading-twist contributions, one should expect that some higher-twist effects might taint the values extracted from data.
Nevertheless, this procedure will provide an order of magnitude of the impact of kinematic-power corrections.
For this analysis, we used the numerical implementation of the GK model found in PARTONS \cite{Berthou:2015oaw} and use the Apfel++ library \cite{Bertone:2013vaa,Bertone:2016lga,Bertone:2017gds,Bertone:2022frx} to perform the leading order evolution from $4\ \textrm{GeV}^2$ to $2\ \textrm{GeV}^2$. 

\subsubsection{Model of the $D$-term}

The $D$-term extracted in Ref.~\cite{Dutrieux:2024bgc} was based on a Gegenbauer decomposition, following Eq.~\eqref{eq:GegenbauerDecomposition}, and thus only the $d_{2n+1}(t)$ coefficients have to be extracted.
Note that the $t$-dependence is modelled by :
\begin{align}
  \label{eq:tModel}
  d_{2n+1}(t) = \frac{ d_{2n+1}(0)}{\left(1-\frac{t}{\Lambda_{2n+1}^2}\right)^3}.
\end{align}
where the power-law behaviour is fixed and not extracted.
In practice, because of the deconvolution problem of the $D$-term already mentioned, only $d_1(t)$ is extracted with a large uncertainty.
We assume all higher Gegenbauer modes to be zero.
Ref. \cite{Dutrieux:2024bgc} reports $d_1(0) = -0.7 \pm 1.3$.
In the following we will stick to the central value of $-0.7$ in order to obtain an order of magnitude of the next-to-leading power corrections impact.
For completeness, we highlight that $\Lambda_1 = 0.8\mbox{ GeV}$.

\subsubsection{Ratio computation}

The DVCS subtraction constant $\mathcal{S}$ is obtained by summing the different flavour contributions, weighted by their electric charge:
\begin{align}
  \label{eq:Sdef}
  \mathcal{S}(t,Q^2) = \sum_q e_q^2 h_0^{++,q}.
\end{align}
Using Eq.~\eqref{eq:Dscr_final}, one can readily get the pure leading-twist contribution:
\begin{align}
  \label{eq:SLT}
  \mathcal{S}_{LT}(t,Q^2) \approx 4\sum_q e_q^2 d_1^q(t,Q^2) \approx 4d_1(t,Q^2) \sum_q e_q^2 ,
\end{align}
where we have used the fact that a flavour-degenerate value of $d_1$ was extracted in Ref.~\cite{Dutrieux:2024bgc}.
Note however that, for computing the part corresponding the DD $F$ and $K$, we have kept the flavour dependence modelled by Goloskokov and Kroll. 

\begin{figure}[t]
  \centering
  \includegraphics[width=0.5\textwidth]{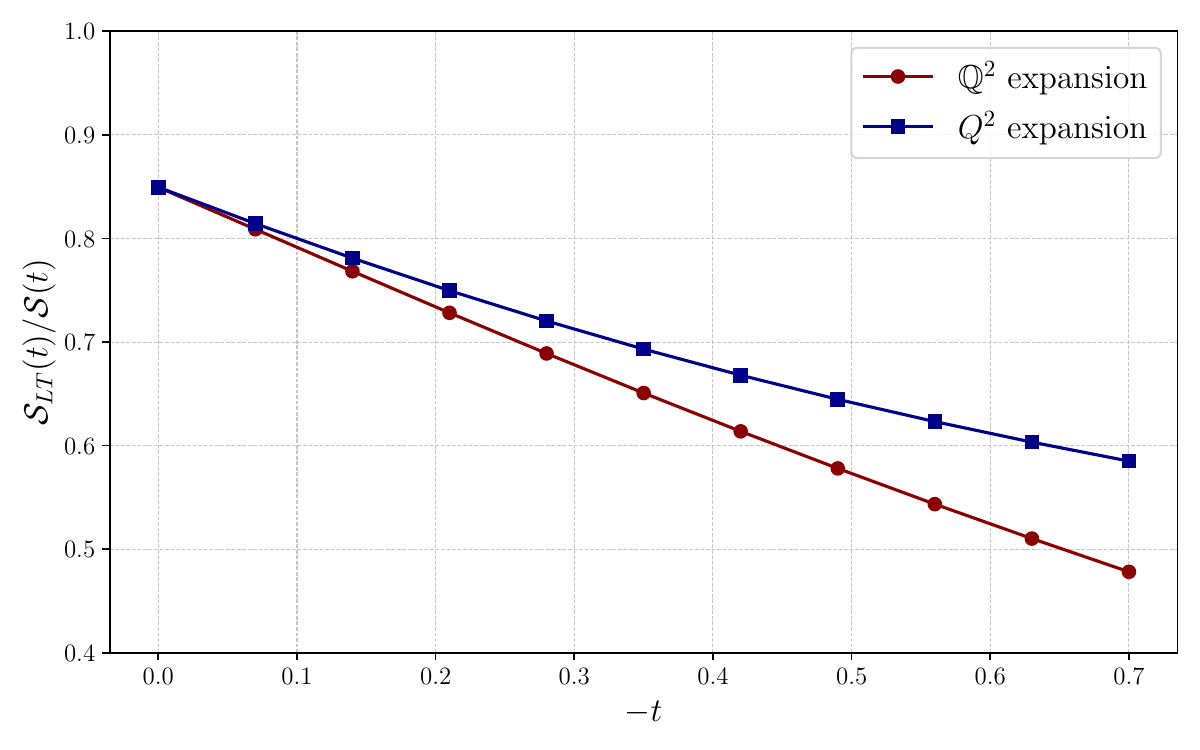}
  \caption{Ratio $\mathcal{S}_{LT} / \mathcal{S}$ as a function of $t$ for $Q^2 = 2\mbox{ GeV}^2$. In red, $\mathcal{S}$ is expanded as power of $1/\scale^2$, while in blue it is in terms of power of $1/Q^2$. The difference is expected to be twist-6 but it is already visible for $|t|$ as small as $0.3\mbox{ GeV}^2$. The dots and squares are the actual computed points, the line is only an interpolation used as a guide.}
  \label{fig:PlotRatio}
\end{figure}

The ratio of the pure leading-twist subtraction constant to the total one is shown on Fig.~\ref{fig:PlotRatio}.
This figure highlight the important impact of the higher-twist corrections, going from 15\% to 50\% on the kinematic range selected.
We also illustrate the impact of choosing $Q^2$ or $\scale^2$ as the typical power-expansion scale.
Above $|t|> 0.3\ \textrm{GeV}^2$ the difference becomes noticeable highlighting the potential need for a twist-6 correction.


\section{conclusion}
\label{sec:conclusion}

In this work, we have generalised the dispersion relations of DVCS to include higher-twist kinematic power corrections, both for spin-0 and spin-$1/2$ targets.
The results are two-fold.
First, we prove that the expression for the $n$‑subtracted dispersion relations is the same as at leading twist (see Eq.~\eqref{eq:DR_Standard}).
This follows from the fact that the imaginary part of the CFF is generated solely by the DGLAP region of the GPD—a result that was, at least to us, unexpected.
The second important point is the modification of the coefficient function; the latter being itself connected to GPDs and DDs. 
These modifications are such that an additional term is introduced and comes with a dependence in the full double distributions $F$ and $K$, and cannot be considered  suppressed, especially for JLab kinematics.
It calls into question the common thought that DVCS dispersion relations allow one to bypass the extraction of GPDs to get access to the pressure and shear forces of quarks within the nucleon. This issue is discussed in a companion paper \cite{Martinez-Fernandez:2025jvk}.
Note that our results for the one-time subtraction constant for the spin-1/2 target~\eqref{eq:h0++_tw4_full_spin1/2} expressed in terms of Double Distribution is in agreement with a previous proof obtained in Ref.~\cite{Braun:2014sta} in terms of GPDs and recalled here in Eq.~\eqref{eq:BMMPResult}.
  We also derived the kinematic twist-four corrections to higher subtracted relations in Eqs.~\eqref{eq:hn++_DD} and \eqref{eq:hn++_tw4_full_spin1/2} for spin-0 and spin-1/2, respectively.
  Finally, Eq.~\eqref{eq:Dscr_final} provides the first calculation of the higher-twist convolution with the Polyakov-Weiss $D$-term.

That being written, assuming that the additional term is taken into account by some procedure, we also investigated the impact of the kinematic corrections to the deconvolution problem of the $D$-term.
To do so, we stay to leading order in $\alpha_S$, two Gegenbauer modes, and studied how the shadow $D$-term is impacted by the $t/\scale^2$ corrections.
If the coefficient becomes indeed dependent of the mode $n$ considered, this dependence is suppressed quadratically in $n$, which precludes any help for deconvoluting modes beyond the first few ones. 
And if one is restricted to the helicity conserving amplitudes, then this mode dependence is a small perturbation of the mode independent, leading-twist part.
So that, if the situation may improve with respect to the pure LT, we do not expect a significant effect compared to the one provided by evolution (which is itself already small).

This leads to the considerations of future work.
On the one hand, with a large contribution from the \mbox{twist-4} nucleon mass correction at JLab kinematics, we wonder what happens at kinematic \mbox{twist-6}.
On the other hand, we believe that studying the photon helicity flip amplitude, $\mathcal{H}^{+-}$, would be of great interest.
Indeed, at leading order, there is no pure leading-twist contribution, which may allow one to disentangle between the two first Gegenbauer mode.
However, one would also need to assess the impact of gluon ``transversity'' GPDs, that contribute to  $\mathcal{H}^{+-}$ at NLO.

With all these points in mind, we believe that more theoretical and phenomenological work is necessary before we can extract reliable distributions of pressure and shear forces within the nucleon from experimental data.

\begin{acknowledgments}
  The authors are grateful to V. Braun for highlighting the previous results obtained in the literature.
  The authors would also like to thank V. Bertone, H. Dutrieux, C. Lorcé, B. Pire and P. Sznajder for stimulating discussion.
  This research was funded in part by l’Agence Nationale de la Recherche (ANR), project ANR-23-CE31-0019.
  For the purpose of open access, the author has applied a CC-BY public copyright licence to any Author Accepted Manuscript (AAM) version arising from this submission.
  This work was made possible by Institut Pascal at Université Paris-Saclay with the support of the program “Investissements d’avenir” ANR-11-IDEX-0003-01.
\end{acknowledgments}

\appendix

\section{Coefficient functions}
\label{app:coefficientfunction}
\subsection{Leading-twist coefficient function}

We start by providing the leading-twist coefficient function for the helicity conserving amplitude taken from Ref.~\cite{Moutarde:2013qs} up to NLO in $\alpha_s$:
\begin{align}
    C_0(x/\xi) =&\ \frac{-1}{x/\xi-1+i0} \,, \label{eq:C0} \\
    C_1(x/\xi) =&\ \frac{\alpha_s C_F}{4\pi} \frac{1}{x/\xi+1-i0}
    \bigg[
                  9 -3\frac{x+\xi}{x-\xi}\Ln{\frac{x+\xi}{2\xi}-i0} \nonumber \\
                &     -\ln^2\left(\frac{x+\xi}{2\xi}-i0 \right)\bigg]\,, \\
    C_{\rm coll} (x/\xi) =&\ \frac{\alpha_s C_F}{4\pi}
    \frac{1}{x/\xi+1-i0}
    \bigg[
    -3-2\Ln{\frac{x+\xi}{2\xi}-i0} \bigg]  
    \,.
\end{align}
Defining the antisymmetric coefficients
\begin{align}
    C^{(+)}_0(x/\xi) = &\ C_0(x/\xi) -(x/\xi\to -x/\xi)\,,\\
    C^{(+)}_1(x/\xi) = &\ C_1(x/\xi) -(x/\xi\to -x/\xi)\,,\\
    C^{(+)}_{\rm coll} (x/\xi) = &\ C_{\rm coll} (x/\xi) -(x/\xi\to -x/\xi)\,,
\end{align}
they combine to give the full LT coefficient function
\begin{align}
  \label{eq:CLT}
  C^{(+)}_{\rm LT}\left(\frac{x}{\xi}\right) =  & \left[ C^{(+)}_0\left(\frac{x}{\xi}\right) + C^{(+)}_1\left(\frac{x}{\xi}\right) \right. \nonumber \\
    & +\left.  \Ln{\frac{Q^2}{\mu_{\rm F}^2}}C^{(+)}_{\rm coll}\left(\frac{x}{\xi}\right) \right] + O(\alpha_s^2) \,.
\end{align}

\subsection{Twist-4 Scalar Coefficient Function for $\cffH^{++}$}

To kinematic twist-4 accuracy, the CFF $\cffH^{++}$ reads\footnote{$\amp^{++}_{\textrm{\cite{Martinez-Fernandez:2025gub}}} = \cffH^{++}$~at 0th-order in $\alpha_s$.}~\cite{Braun:2022qly,Braun:2025xlp,Martinez-Fernandez:2025gub}
\begin{align}
    \cffH^{++} = & \int_{-1}^1 dx\ \frac{1}{\xi}\Bigg\{ \left(1-\frac{t}{2\scale^2}\right)C^{(+)}_0(x/\xi)H \nonumber\\
    & + \frac{t}{\scale^2} \left[ \wpbbiii^{(+)}(x/\xi)-\frac{\mathcal{L}^{(+)}(x/\xi)}{2} \right]H \nonumber\\
    & -\frac{t}{2\scale^2}\xi\dxi\left[\left( \wpbbiii^{(+)}(x/\xi) - \mathcal{L}^{(+)}(x/\xi) \right)H\right] \nonumber\\
    & + \frac{\xi^2\bp_\perp^2}{\scale^2}\xi^2\dxi^2\left[\left( \wpbbiii^{(+)}(x/\xi) - \mathcal{L}^{(+)}(x/\xi) \right)H\right]
      \Bigg\} \nonumber \\
  & + O(\alpha_s,\,\textrm{tw-6},\,\alpha_s\cdot\textrm{tw-4})\,
\end{align}
where we truncated $C_{\rm LT}$ to order $(\alpha_S)^0$ following eq. \eqref{eq:CLT}.
Here, we introduced the notation:
\begin{align}
  \label{eq:wpbbiii}
  \wpbbiii^{(+)}(x/\xi) & = \wpbbiii(x/\xi) - (x/\xi\to -x/\xi)\,,\\
  \label{eq:calL}
  \mathcal{L}^{(+)}(x/\xi) & = \mathcal{L}(x/\xi)  - (x/\xi \to -x/\xi)\,,
\end{align}
upon
\begin{align}
    \wpbbiii(x/\xi) & = \frac{-2}{x/\xi+1}\Ln{\frac{x/\xi-1+i0}{-2+i0}} \,,  \label{wpbbiii} \\
    \mathcal{L}(x/\xi) & = \frac{4}{x/\xi-1}\left[ \Li{2}{\frac{x/\xi+1}{2-i0}} - \Li{2}{1} \right]\,. \label{calL}
\end{align}

\subsection{Twist-4 spin-1/2 coefficient function for $\cffH^{++}$ and for $\cffE^{++}$}\label{app:spin-1/2_kernels}

The hard coefficient functions of a spin-1/2 particle at LO and up to kinematic twist-4 accuracy where detailed in Eq.~\eqref{eq:BraunYiMansahovTw4Kernels}, see main text. All in all, these kernels provide an alternative but equivalent formulation to the spin-0 basis of functions for computing the kinematic corrections. However, one must notice that the final convolutions giving rise to the $\cffH^{++}$ of a spin-0 and spin-1/2 targets are not fully equivalent, vid.~Eq.~\eqref{eq:cffH_spin1/2_decomposition}. This can be understood by the works of A.~V.~Belitsky \& D.~M\"uller~\cite{Belitsky:2000vx}, and V.~M.~Braun \& A.~N.~Manashov~\cite{Braun:2011dg} where it is manifest that the kinematic twist corrections stem from higher-order diagrams involving gluon exchanges between the active quark in the scattering and the spectator structure of the hadron. In fact, it is only at leading twist that all those exchanges can be fully resummed into the well-known Wilson links that guarantee the gauge-invariant properties of GPDs.

Conversely, NLO corrections in perturbation theory consisting of loop diagrams involving self-energy and quark-photon vertex corrections that carry information on the spin and nature of said active quark and not on the characteristics of the hadron from where this parton has been originated. Consequently, at NLO-LT accuracy, there is no difference in the hard kernel $C_{\textrm{LT}}$ between targets of different spins.


\section{Gegenbauer Polynomials}
\label{app:gegenbauer}
The Gegenbauer polynomials $\gegen{n}{\lambda}(x)$ are a specific case of Jacobi polynomials of degree $n$, such that they are othogonal for a certain weight of the type $(1-x^2)^{\lambda -\frac{1}{2}}$.
For instance, considering $\lambda=\frac{3}{2}$, one gets the first terms as
\begin{align}
    \gegen{1}{3/2}(\alpha) &= 3\alpha \,,\\
    \gegen{3}{3/2}(\alpha) &= \frac{5}{2}\alpha(7\alpha^2-3) \,,
\end{align}
and an orthogonality condition provided by:
\begin{align}
  &\int_{-1}^1 d\alpha\ (1-\alpha^2)\gegen{n}{3/2}(\alpha)\gegen{m}{3/2}(\alpha)\nonumber \\
  = &\delta_{n,m}\frac{\pi\Gamma(n+3)}{n!4(n+3/2)\left[ \Gamma(3/2) \right]^2} \nonumber\\
     = &\delta_{n,m}\frac{2(n+3)(n+2)(n+1)}{2n+3} \,.
\end{align}
These polynomials satisfy two properties that will be of use later on:
\begin{align}
    \gegen{N}{\lambda}(-\alpha) & = (-1)^N\gegen{N}{\lambda}(\alpha) \,, \label{prop::parity} \\
  2(n+\lambda)\gegen{n}{\lambda}(\alpha) & = \frac{d}{d\alpha}\left[ \gegen{n+1}{\lambda}(\alpha) - \gegen{n-1}{\lambda}(\alpha) \right]\nonumber \\
  & = 2\lambda\left[\gegen{n}{\lambda+1}(\alpha) - \gegen{n-2}{\lambda+1}(\alpha)\right] \,, \label{prop::difference_gegenbauers} \\
    \sum_{N=0}^\infty \gegen{N}{\lambda}(\alpha) \tau^N & = \frac{1}{(1-2\alpha \tau + \tau^2)^\lambda}\,,\quad |\tau|\leq 1\,.  \label{prop::generating_functional}
\end{align}

\bibliographystyle{unsrt}
\bibliography{../Bibliography}

\end{document}